\documentclass[numberedappendix]{emulateapj}

\usepackage{apjfonts,subfigure}

\def\LCDM{$\Lambda\mbox{CDM}$}
\def\Msun{\hbox{$\rm\thinspace M_{\odot}$}}
\def\kpc{{\rm\thinspace kpc}}
\def\Mpc{{\rm\thinspace Mpc}}

\def\Gyr{{\rm\thinspace Gyr}}

\shorttitle{The Millennium Simulation halo merger rate}
\shortauthors{Genel et al.}

\begin{document}

\title{The halo merger rate in the Millennium Simulation and implications for observed galaxy merger fractions}

\author{Shy Genel\altaffilmark{1}, Reinhard Genzel\altaffilmark{1,2}, Nicolas Bouch{\'e}\altaffilmark{1}, Thorsten Naab\altaffilmark{3}, Amiel Sternberg\altaffilmark{4}}

\altaffiltext{1}{Max Planck Institut f\" ur extraterrestrische Physik, Giessenbachstrasse, D-85748 Garching, Germany; shy@mpe.mpg.de; genzel@mpe.mpg.de; nbouche@mpe.mpg.de.}
\altaffiltext{2}{Department of Physics, Le Conte Hall, University of California, Berkeley, CA 94720.}
\altaffiltext{3}{Universit\"ats-Sternwarte M\"unchen, Scheinerstr.\ 1, D-81679 M\"unchen, Germany; naab@usm.uni-muenchen.de.}
\altaffiltext{4}{School of Physics and Astronomy, Tel Aviv University, Tel Aviv 69978, Israel; amiel@wise.tau.ac.il.}

\slugcomment{Accepted for publication in The Astrophysical Journal}

\begin{abstract}
We have developed a new method to extract halo merger rates from the Millennium Simulation. First, by removing superfluous mergers that are artifacts of the standard Friends-Of-Friends (FOF) halo identification algorithm, we find a lower merger rate compared to previous work. The reductions are more significant at lower redshifts and lower halo masses, and especially for minor mergers. Our new approach results in a better agreement with predictions from the extended Press-Schechter model. Second, we find that the FOF halo finder overestimates the halo mass by up to $50\%$ for halos that are about to merge, which leads to an additional $\approx20\%$ overestimate of the merger rate. Therefore, we define halo masses by including only particles that are gravitationally bound to their FOF groups. We provide new best-fitting parameters for a global formula to account for these improvements. In addition, we extract the merger rate per {\it progenitor} halo, as well as per descendant halo. The merger rate per progenitor halo is the quantity that should be related to observed galaxy merger fractions when they are measured via pair counting. At low mass/redshift the merger rate increases moderately with mass and steeply with redshift. At high enough mass/redshift (for the rarest halos with masses a few times the "knee" of the mass function) these trends break down, and the merger rate per progenitor halo decreases with mass and increases only moderately with redshift. Defining the merger rate per progenitor halo also allows us to quantify the rate at which halos are being accreted onto larger halos, in addition to the minor and major merger rates. We provide an analytic formula that converts any given merger rate per descendant halo into a merger rate per progenitor halo. Finally, we perform a direct comparison between observed merger fractions and the fraction of halos in the Millennium Simulation that have undergone a major merger during the recent dynamical friction time, and find a fair agreement, within the large uncertainties of the observations. Our new halo merger trees are available at http://www.mpe.mpg.de/ir/MillenniumMergerTrees/.
\end{abstract}


\section{Introduction}
\label{s:intro}
In the current paradigm for structure formation, the cold dark matter model \citep{WhiteS_78a,BlumenthalG_84a,DavisM_85a,SpringelV_06a}, galaxy mergers play an important role in galaxy formation and evolution. Galaxy mergers drive gas towards central starbursts (e.g.~\citealp{MihosJ_96a}) and supermassive black holes (e.g.~\citealp{HernquistL_89a}), and transform galactic morphology (e.g.~\citealp{NaabT_03a,BournaudF_05a}). Many parameters, like the gas fraction and morphology of the merging galaxies, as well as their relative orbits and orientations, affect the properties of the merger remnant. One of the most important factors is the mass ratio of the merging galaxies (e.g.~\citealp{NaabT_06b}). Galaxy mergers of all mass ratios are frequent in a \LCDM{ }Universe (e.g.~\citealp{StewartK_08a}), but special importance is given to major mergers, usually considered as those with mass ratios less than $\approx3:1$. These are thought to play a significant role in the buildup of the red sequence \citep{ToomreA_77a,HopkinsP_07b} by transforming blue star-forming late type galaxies into red passive early type galaxies.

Many observational studies have been carried out in recent years to investigate the fraction of galaxies that show signs of major merger activity as a function of mass, luminosity and redshift. Observationally, only merger fractions can be obtained, and in order to transform them into merger rates, which can be directly compared with theoretical models, the time scale of the observed events must be estimated (e.g.~\citealp{PattonD_00a}). Two principal approaches are used to observe galaxy merger fractions. One is pair counting (e.g.~\citealp{PattonD_97a,LeFevreO_00a,BellE_06b,RyanR_07a,LinL_08a}), i.e.~identifying galaxies separated from one another by less that typically $\approx20\kpc$. This method probes the pre-merger stage and is therefore a ``progenitor galaxy'' merger fraction \citep{DeProprisR_07a}. The second approach is identification of mergers through morphological signatures such as asymmetry and tidal tails (e.g.~\citealp{LeFevreO_00a,ConseliceC_03a,LotzJ_08a}). This method aims at identifying mergers in their relatively late stages, and is therefore a ``descendant galaxy'' merger fraction. For each of the principal approaches different authors use different methods, as well as different selection criteria, which create systematic effects that are not easily comparable. Moreover, the intrinsic uncertainties of each method, and the effect of cosmic variance (e.g.~\citealp{ConseliceC_08a}) contribute further to the large scatter between the obtained results.

\citet{DeProprisR_07a} and \citet{ConseliceC_06a} have noticed that the merger fractions obtained with the two principal methods should be carefully defined, and cannot be directly compared to each other, as they are actually different quantities. Both authors proposed (different) simple conversions between the two quantities. Nevertheless, both disregarded the fact that the mass difference between the progenitors and the descendant is up to a factor of $2$ (in equal-mass binary mergers). That may result in very large differences in the number densities of progenitors and of descendants. Since the measured merger {\it fraction} implicitly includes the information of the {\it number} of objects, different number densities are implicitly included in the merger rate or merger fraction per progenitor galaxy versus per descendant galaxy. \citet{BellE_06a} and \citet{LotzJ_08a} realised this, and have taken the different number densities into account when comparing merger fractions obtained with different methods. We show in this paper that the different definitions of the merger rate, per descendant or per progenitor halo, lead to significantly different results, particularly for major mergers, high redshift mergers, and high mass mergers.

Galaxy mergers follow their dark matter halo mergers, but the connection between the two is not straightforward. When two dark matter halos merge, the orbital angular momentum is transferred into internal degrees of freedom, while the more concentrated galaxies are at first not much affected. The galaxies lose relative angular momentum due to dynamical friction, and they start merging as well when they are more tightly bound \citep{BarnesJ_92a}. All of the baryonic physics involved in galaxy mergers makes it difficult theoretically to quantify the galaxy merger rate reliably (P.~F.~Hopkins et al.~2009, in preparation). Nevertheless, a first step towards quantifying the galaxy merger rate would be to understand the dark matter halo merger rate, which can be studied more robustly. 

\citet{LaceyC_93a} have estimated the halo merger rate based on the extended Press-Schechter (EPS) model \citep{PressW_74a,BondJ_91a,BowerR_91a}. \citet{NeisteinE_08b} and \citet{ZhangJ_08b} have recently constructed EPS-based approximations that are self-consistent, and \citet{ZhangJ_08a} have done so for the ellipsoidal collapse model. All these models differ by factors of a few tenths up to a few. N-body simulations have only recently been used to study the halo merger rate with a large dynamical range (\citealp{DOnghiaE_08a}; \citealp{StewartK_08b}; \citealp{FakhouriO_07a}, hereafter FM08). FM08 have investigated this problem based on the large Millennium Simulation \citep{SpringelV_05a}. They have found that the dark matter halo merger rate has an almost universal form that can be separated into its dependencies on mass ratio, descendant mass and redshift. Nevertheless, the analysis of N-body simulations is also subject to uncertainties, in particular in identifying halos. FM08 present 3 ways to analyse the simulation, which differ from one another by $\approx25\%$, and by a more significant amount compared to the \citet{LaceyC_93a} analytical approximation. All of these investigations have quantified the halo merger rate per {\it descendant} halo. Recent work has also used N-body simulations to study mergers of subhalos \citep{AnguloR_08a,WetzelA_08a}.

In this paper we extract the dark matter halo merger rate from the Millennium Simulation using a new method for identifying halos and mergers, as well as a new definition of the merger rate. In \S\ref{s:millennium} we review the Millennium Simulation and its post-processing. In \S\ref{s:new_trees} we describe how we create merger trees that are free of artificial effects not considered previously. In \S\ref{s:rates} we define the merger rate per descendant halo, as well as the merger rate per progenitor halo, which is related to observed galaxy merger fractions when they are measured via pair counting. In \S\ref{s:results} we describe our results and in \S\ref{s:comparison} we compare them to previous work. In \S\ref{s:merger_fractions} we calculate the halo merger fraction in the Millennium Simulation and compare to observations. In \S\ref{s:summary} we discuss our results and their relevance to observations, and summarise.

\section{Extracting the merger rate from the Millennium Simulation}
\label{s:find_merger}

\subsection{The Millennium Simulation and its merger trees}
\label{s:millennium}
The Millennium Simulation is a cosmological N-body simulation following $2160^3$ dark matter particles, each of mass $8.6\times10^8h^{-1}\Msun$, in a box of $500h^{-1}\Mpc$ on a side, with 64 generated output times (``snapshots'') from $z=127$ to $z=0$. The cosmology is set to \LCDM{ }with $\Omega_m=0.25$, $\Omega_{\Lambda}=0.75$, $\Omega_b=0.045$, $h=0.73$, $n=1$ and $\sigma_8=0.9$, which we will adopt throughout this paper.

In the Millennium Simulation there are two levels of structure identification. First, the Friends-Of-Friends (FOF) algorithm (\citealp{DavisM_85a}; with a linking length $b=0.2$ of the mean particle separation) creates a catalogue of FOF groups in every snapshot. In the limit of a large number of particles, FOF groups enclose the particles within isodensity contours of $\approx b^{-3}$ times the mean matter density \citep{FrenkC_88a,LaceyC_94a,JenkinsA_01a}, and their mean densities correspond approximately to the overdensities of virialised halos expected from the spherical collapse model \citep{LaceyC_94a}. Thus, FOF groups are considered to represent dark matter halos. However, substructure as traced by local density maxima are not distinguishable within the FOF groups. For this purpose, the algorithm SUBFIND \citep{SpringelV_01} identifies substructures in each FOF group, by finding gravitationally self-bound groups of particles around maxima in their smoothed density field. Thus, each FOF group contains at least one subhalo, and the subhalo with the "most massive history" is chosen to be the {\it main} subhalo \citep{DeLuciaG_07a}.

The Millennium merger trees\footnote{Structure catalogues and the derived merger trees have been made public by the Virgo Consortium: http://www.mpa-garching.mpg.de/millennium.} are constructed from the subhalos by finding a single descendant for each subhalo at the following snapshot. The FOF groups themselves play no role in constructing the merger trees. Nevertheless, if the halo merger rate is to be studied, then new merger trees, in which each node is a halo rather than a subhalo, must be built.

\subsection{Constructing new halo merger trees}
\label{s:new_trees}
We build new halo-based, that is FOF group-based, merger trees by defining one descendant for each FOF group. The main subhalo in each FOF group is identified and followed to its subhalo descendant using the original subhalo trees. Then, the FOF group to which that subhalo descendant belongs is defined as the FOF group descendant of the FOF group in question.

\begin{figure*}[tbp]
\centering
\subfigure[]{
          \label{f:distance_during_merger}
          \includegraphics[]{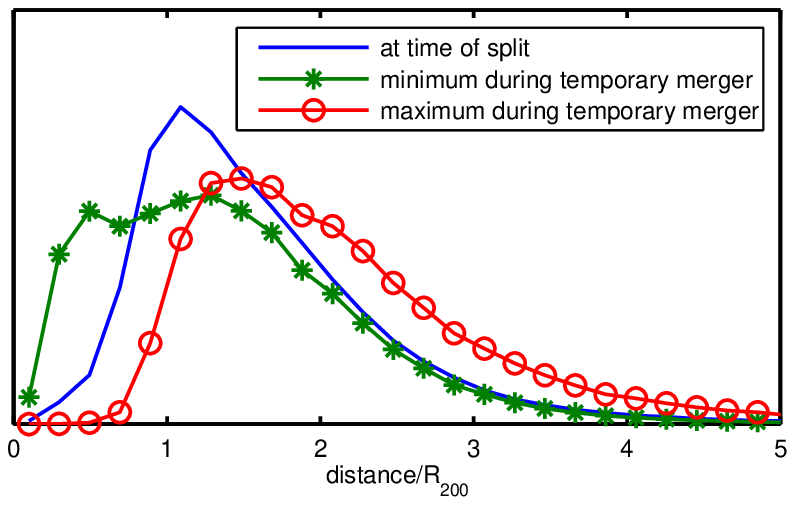}}
\subfigure[]{
          \label{f:relative_durations}
          \includegraphics[]{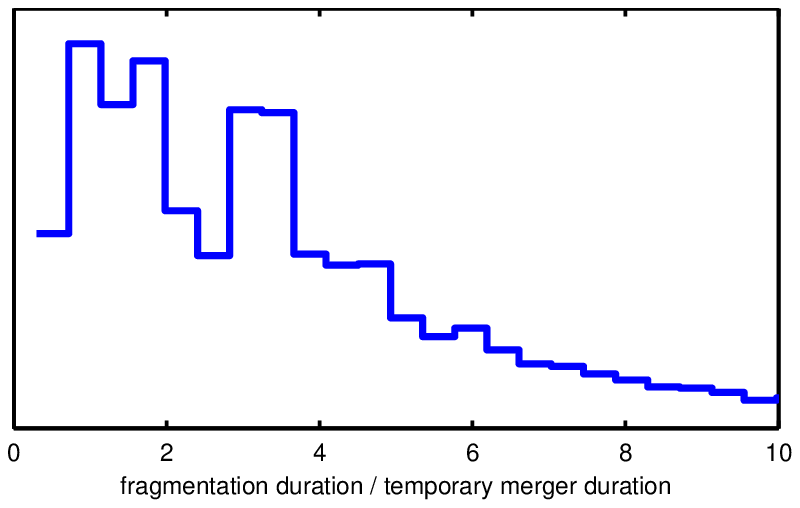}}
\caption{{\it Left:} The distribution of relative distances (with respect to $R_{200}$ at each snapshot) between subhalos that only temporarily belong to the same FOF group. The solid blue curve shows the distribution at the last snapshot of the temporary merger, the green curve with asterisks shows the minimum relative distances during the temporary merger, and in red with circles is the distribution of maximum distances. These distributions are shifted slightly towards lower (higher) values at lower (higher) redshifts and for more minor (major) mergers. {\it Right:} The distribution of the ratio between the fragmentation duration (the time between the fragmentation itself and the subsequent re-merger) and the first, temporary, merger that precedes the fragmentation, for cases where the fragmentation lasts more than three snapshots. The fragmentation duration is typically much longer (mean $\approx4$, median $\approx3$) than the temporary merger duration, indicating that the temporary merger is usually only an artifact of the FOF algorithm.}
\vspace{0.5cm}
\label{f:splitting_motivation}
\end{figure*}

\begin{table*}[tbp]
\caption{Merger scenarios, their abundance, and the different counting by different algorithms}
\label{t:counting_mergers}
\centering          
\begin{tabular}{ | c | c | c | c | c | c | c | c | } 
\hline\hline 
first merger duration in snapshots  & $\infty$ & $\leq3$    & $\leq3$    & $\leq3$    & $>3$        & $>3$       & $>3$ \\
\hline
fragmentation duration in snapshots & -        & $\leq3$    & $>3$       & $\infty$   & $\leq3$     & $>3$       & $\infty$ \\ 
\hline
abundance                           & $78\%$   & $7.6\%$    & $4.4\%$    & $3.5\%$    & $2.5\%$     & $1.9\%$    & $2.1\%$ \\ 
\hline
snipping                        & 1        & \textbf{2} & \textbf{2} & \textbf{1} & \textbf{2}  & \textbf{2} & \textbf{1} \\ 
stitching-3                     & 1        & 1          & \textbf{2} & \textbf{1} & 1           & \textbf{2} & \textbf{1} \\ 
stitching-$\infty$              & 1        & 1          & 1          & \textbf{1} & 1           & 1          & \textbf{1} \\ 
splitting-3                     & 1        & 1          & 1          & 0          & \textbf{2}  & \textbf{2} & \textbf{1} \\
splitting                       & 1        & 1          & 1          & 0          & 1           & 1          & 0 \\ 
\hline                  
\end{tabular}
\tablecomments{In bold font are scenarios for which a certain algorithm over-counts the number of mergers. For details see \S\ref{s:new_trees}. A merger duration of $\infty$ snapshots denotes a merger with no fragmentation, and a fragmentation duration of  $\infty$ snapshots means that the fragments never re-merge by $z=0$. The abundances of the different scenarios are calculated using all the mergers in the Millennium Simulation.}
\vspace{0.5cm}
\end{table*}

In practice, FOF groups not only merge, but may also split. This is not compatible with the simplified notion of hierarchical build-up and indeed not described at all by an analytical model such as EPS. Yet, this phenomenon is robust in numerical N-body simulations. A FOF group is described as ``split'' when a {\it subhalo} belonging to it at a given snapshot is no longer part of it at a later snapshot.

Those subhalos that were split out of their FOF groups can be classified according to several criteria, describing their past and future histories: where they were created, what structure they belong to right after the split, and where they end up at later times. Almost all subhalos are not {\it formed} inside their host FOF groups, but originate from previous mergers in which they were accreted onto them. Thus, most split subhalos are part of mergers that started in the past but were cut by the split, before the subhalos merged. The location of the split subhalo in the immediate snapshot after the split can be either as an independent FOF group or as a subhalo in another FOF group. After the split, the split subhalo may re-merge with the FOF group from which it was split, or it may not do so at all.

A very typical case is that of two FOF groups that merge, with one becoming a subhalo of the other, which then later split again into two distinct FOF groups. If they never re-merge, it is clear that a ``merger'' interpretation of such an event would not be appropriate. Most likely, a temporary bridge of particles caused the FOF algorithm to identify them as one FOF group \citep{WhiteM_01a,LukicZ_08a}. If they eventually do re-merge after a few snapshots, the first merging and splitting merely shows that they were relatively close and that some interaction took place. However, the actual point of merging should correspond to the merger event after which there is no more splitting. To support that interpretation, we show in Figure \ref{f:distance_during_merger} that in $65\%$ ($85\%$) of temporary mergers the distance between the subhalos never decreases below $R_{200}$ ($R_{200}/2$) of the joint FOF group. The undesired implications of the way splits appear in the original merger trees are artificial changes in the halo mass function and merger counts. As shown by FM08 and as will be shown later on in this paper, different algorithms that deal with this phenomenon change the halo mass function only slightly, but have a pronounced effect on the merger rate.

It is common not to treat such fragmentation cases in any special way, thereby allowing the split fragments to have no progenitors (e.g.~FM08's "snipping" method). In that case, the split fragment may re-merge (even several times) with the same halo. One approach of dealing with this issue (e.g.~\citealp{HellyJ_03a,HarkerG_06a}) is via different algorithms for splitting FOF groups that are artificial combinations of several halos. Another approach was taken by FM08, who chose in their fiducial "stitching-3" method to merge back the split fragments when a future merger takes place within three subsequent snapshots after the fragmentation. They also presented a method, "stitching-$\infty$", where subhalos are never allowed to split out of their FOF groups. In the present paper we eliminate the occurrence of halo fragmentation in our new trees, in a way similar to \citet{HellyJ_03a}. We do so by identifying any FOF group at redshift $z_p$ that contains at least a pair of subhalos that at some lower redshift $z_f<z_p$ do not belong to the same FOF group. Such a FOF group is split by our algorithm into several fragments in the following way. Subhalos which belong to different FOF groups at $z_f$ will belong to different fragments at $z_p$ as well, while subhalos that do stay together at $z_f$ will not be separated at $z_p$. Any new fragment our algorithm creates, as well as any untouched FOF group, is considered hereafter simply as a "halo". \citet{FakhouriO_08b} (hereafter FM09) have suggested a variant of this algorithm, "splitting-3", that splits the progenitors of fragmenting FOF groups only 3 snapshots backwards.

We identify two possible advantages of our "splitting" method over "snipping", "stitching" and "splitting-3" algorithms. The first has to do with double counting of mergers. Table \ref{t:counting_mergers} presents the possible scenarios for binary merging and fragmentation events, and the number of mergers that each of the methods presented above counts for those scenarios. Printed in a bold font are cases where a method counts more mergers than appropriate. Only our "splitting" method never counts artificial mergers as real ones. Moreover, all other methods except from "stitching-$\infty$" leave spurious fragmentations in the tree, which may be mistakenly interpreted as extremely high (positive and negative) "smooth" changes in the halo mass, not associated with any merger. We note, however, that any algorithm that does not double-count mergers, e.g.~a combination of "stitching-3" for short fragmentations and "splitting" for long fragmentations, is valid from this perspective.

The second advantage relates to the timing of the merger, and makes us always prefer "splitting" over a combination of "splitting" and "stitching". In cases described by the second column of Table \ref{t:counting_mergers}, where both the first, temporary, merger and the fragmentation duration are shorter than three snapshots and comparable to each other, there's no significant difference between "stitching" and "splitting". In such cases it is difficult to determine conclusively what the "correct" time is, i.e.~that of the first or the second merger (but note that Figure \ref{f:distance_during_merger}, as mentioned earlier, suggests that the later merger is usually the more physical choice). Nevertheless, when the time spent between the fragmentation and the re-merging is larger than 3 snapshots, the different algorithms perform very differently: "stitching-$\infty$" counts it at the time of the first merger, while our method counts it at the time of the re-merger. As Figure \ref{f:relative_durations} shows, in those cases when the fragmentation phase lasts more than three snapshots, it is mostly a few times longer than the first, temporary, merger. This further suggests that the point of re-merging should be usually considered as the more physically appropriate time of merging.

\begin{figure}[tbp]
\centering
\includegraphics[]{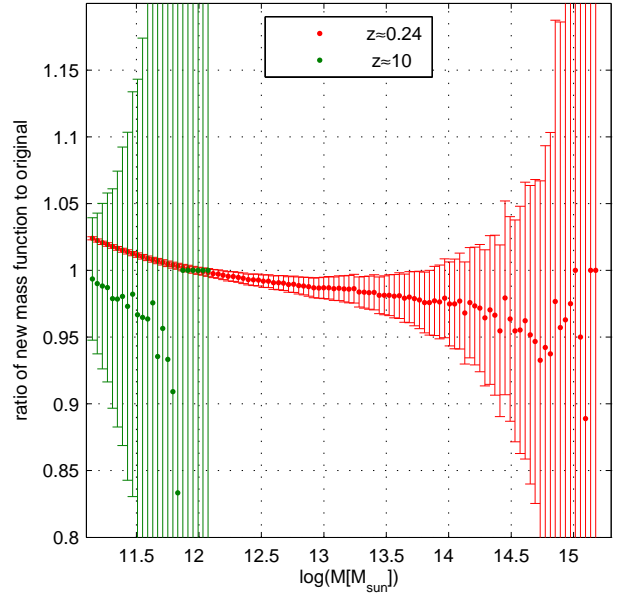}
\caption{The ratio between the halo mass function after applying our splitting algorithm and the original mass function. The difference does not exceed $\approx15\%$, and where it is statistically significant (according to Poisson errors, which are shown by the error bars) it does not exceed $\approx3\%$, for $M>1.5\times10^{11}\Msun$ at all redshifts.}
\vspace{0.5cm}
\label{f:mass_function}
\end{figure}

Our splitting algorithm changes the halo mass function $n(>M,z)$ only slightly. In our new halo catalogues there are naturally fewer massive halos and more small halos, compared to the original FOF group catalogues. Figure \ref{f:mass_function} shows that the change does not exceed $\approx15\%$, while where the change is statistically significant (according to Poisson errors) it does not exceed $\approx3\%$, for $M>1.5\times10^{11}\Msun$ at all redshifts. This difference is smaller than the uncertainty due to e.g.~different identification methods for halos or different simulation codes (e.g.~\citealp{LukicZ_07a,HeitmannK_07a}).

Our method implicitly assumes that no splits occur after $z=0$, which is of course physically wrong but technically unavoidable, since the simulation terminates at $z=0$. The typical time it takes the subhalos to disappear after having merged with bigger FOF groups is comparable to the cosmic lookback time at $z\approx0.1$. Therefore, starting from $z\approx0.2$, it becomes impossible to identify a non negligible fraction of the artificially linked FOF groups - those that would have split after $z=0$ had the simulation continued running. In particular, FOF groups at the last snapshot of the simulation are never split by our splitting algorithm, since it is impossible to know which subhalos of theirs would split ``in the future''. Our results are therefore valid only for $z\gtrsim0.2$, and for use at lower redshifts an extrapolation should be used.

We define the mass of a halo as the mass of all the particles gravitationally bound to it, i.e.~the sum of its subhalos' masses. For $\approx80\%(\approx95\%)$ of the halos, the unbound particles, which we don't take into account, are less than $5\%(20\%)$ of the total FOF mass. Also, the halo mass function $n(>M,z)$ changes by just a few percent if also the unbound particles are included. Nevertheless, including the unbound particles (as in e.g.~FM08) has a significant effect on the inferred merger rate. In \S\ref{s:mass_def} we describe this difference, explain its source and justify our choice of not including the unbound particles in the mass of halos.

\subsection{Definitions of the merger rate: per progenitor halo and per descendant halo}
\label{s:rates}
We identify a merger whenever two or more halos at snapshot $s$ have a common descendant at snapshot $s+1$, and use the time/redshift difference between the two snapshots to define the merger rate per unit time/redshift, respectively. We define a merger as a two-body event, in a way that if $n>2$ halos merge, then $n-1$ mergers are recorded, each between the most massive one and one of the others. The possibility that the mergers occur in a different order, i.e.~that some of the smaller progenitors merge with one another before merging together with the most massive progenitor, is sufficiently small so that our results are not strongly affected by it, as shown by FM08.

We derive the merger rate per {\it progenitor} halo per unit time (or redshift) per mass ratio $x$: $\frac{1}{N_{\rm prog-halo}}\frac{dN_{\rm merger}}{dtdx}(x,z,M)$. Our book-keeping is performed as follows. For a merger between two halos of masses $M_1$ and $M_2$ we record one merger at mass $M_1$ with ratio $x=M_1/M_2$ and one merger at mass $M_2$ with ratio $x=M_2/M_1$. In this way, all the mergers of each halo with any other halo, less or more massive, are recorded. There is no double counting in this method, since we are interested in the merger rate {\it per halo} rather than the absolute number of mergers. We attribute a merger to each halo in the pair, and each merger event is counted as $2/2=1$ merger per halo.

\begin{figure*}[btp]
\centering
\subfigure[]{
          \label{f:rate_vs_x_low_z}
          \includegraphics[]{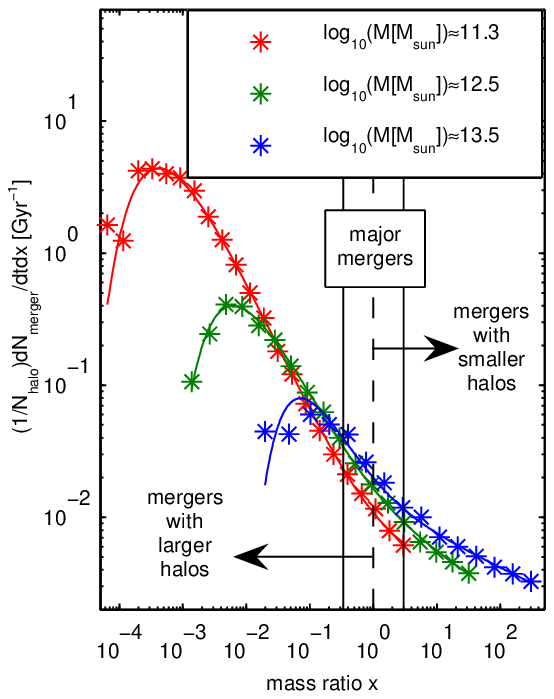}}
\subfigure[]{
          \label{f:rate_vs_x_high_z}
          \includegraphics[]{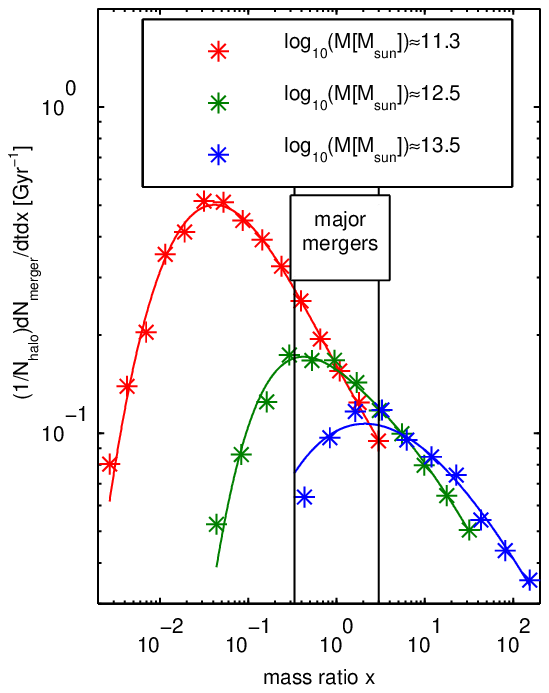}}
\subfigure[]{
          \label{f:rate_vs_x_const_mass}
          \includegraphics[]{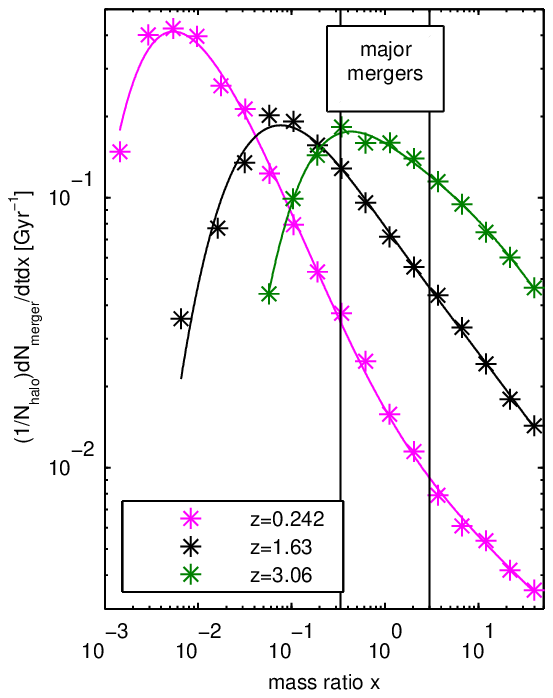}}
\caption{The merger rate $\frac{1}{N_{\rm prog-halo}}\frac{dN_{\rm merger}}{dtdx}(x,z,M)$ as a function of mass ratio $x$. Panel (a) displays the rates for several progenitor masses, at fixed redshift $z\approx0.24$, and Panel (b) shows the same for $z\approx3$. Panel (c) displays the rates at different redshifts, for $10^{12}\Msun<M<2\times10^{12}\Msun$ halos. Asterisks show data based on the simulation, and the solid curves show our fit based on equation (\ref{e:merger_rate}) and Table \ref{t:fits}. To guide the eye, two vertical lines show the range of $x$ within which we define {\it major} mergers. We note that $x$ is limited at large values by the mass resolution limit of the simulation: we show only mergers between $M$ and $M/x>2\times M_{min}=4.72\times10^{10}\Msun$. At small values $x$ is limited by the simulation box size. The bins in $x$ are logarithmically equally-spaced, except for at the left-most side of each curve, where bins were constructed so that they contain no less than 5 mergers.}
\vspace{0.5cm}
\label{f:rates_x}
\end{figure*}

In comparison, for the merger rate per {\it descendant} halo, only one merger is recorded at mass $M_1+M_2$ with ratio $x=M_1/M_2$ (where the indices are defined so that $M_1\geq M_2$). To better understand the difference, let us consider equal-mass mergers ($x=1$). The merger rate derived by using the descendant halo would be one $x=1$ merger per halo at $2M$, while according to our progenitor method it would be one $x=1$ merger per halo at $M$. Therefore, the typical time scale for a halo of mass $M$ to encounter another halo of mass $M$ is the reciprocal of the merger rate {\it per progenitor halo} at mass $M$ and $x=1$. The merger rate per descendant halo at mass $M$ can be quantitatively very different (\S\ref{s:comparison_prog_desc}) and has a different physical meaning. The merger rate per descendant halo gives the time scale on which the population of halos of mass $M$ is {\it created} by equal-mass mergers. It is important to keep in mind the physical meaning of each of the definitions when implementing them to physical problems. For example, as we will discuss in \S\ref{s:summary}, the appropriate quantity to use when considering merger fractions derived via pair counting is the merger rate per progenitor halo.

In the binary merger approximation, which we find to be a good one, both definitions of the merger rate are interchangeable, given that the halo mass function is known (Appendix \ref{s:analytic}). Nevertheless, the merger rate per progenitor halo quantifies also the rate at which halos merge with {\it more massive} halos, i.e.~the range $x<1$, in addition to $x>1$. This is important because the merger rate per descendant halo is given only for the range $x>1$, which represents the rate at which halos of mass $M_1$ merge with (or accrete) less massive halos of mass $M_1/x$. However, the range $x<1$ represents the rate at which halos of mass $M_1$ merge with (or are accreted onto) more massive halos of mass $M_1/x$. To illustrate the importance of this difference, consider the way a major merger between halos of masses $M_1$ and $M_2=M_1/2$ is recorded. When using the descendant halo, one merger with $x=2$ is recorded at mass $3/2M_1=3M_2$, while the fact that the halo with $M_2$ experienced a major merger is not {\it explicitly} accounted for. In contrast, when using the progenitor halos, two mergers are recorded: one with $x=2$ at mass $M_1$ and one with $x=1/2$ at mass $M_2$.

\section{Results}
\label{s:results}
\subsection{The merger rate per progenitor halo as a function of mass ratio}
\label{s:rates_x}
Figure \ref{f:rates_x} shows the merger rate $\frac{1}{N_{\rm prog-halo}}\frac{dN_{\rm merger}}{dtdx}(x,z,M)$ as a function of mass ratio $x$, for a constant redshift at different mass bins (Figure \ref{f:rate_vs_x_low_z} and Figure \ref{f:rate_vs_x_high_z}), as well as for a constant mass bin at different redshifts (Figure \ref{f:rate_vs_x_const_mass}). In all curves, the relation is close to a power law at large $x$. As $x$ decreases, the slope becomes steeper in most curves. At further lower $x$, in all curves, there is a flattening and the sign of the slope changes rapidly as the function starts decreasing towards very small $x$. We fit this shape with the following fitting function:
\begin{eqnarray}
&\frac{1}{N_{\rm prog-halo}}\frac{dN_{\rm merger}}{dtdx}(x,z,M)=\nonumber\\
&Ax^b(1+1/x)^c exp(-(xM_c/M)^d)\Gyr^{-1},
\label{e:merger_rate}
\end{eqnarray}

with $A=A(z,M)$, $b=b(z,M)$, $c=c(z,M)$, $d=d(z,M)$ and $M_c=M_c(z)$.

The $x^b$ term describes the shape of the curves for $x\gg1$, the $(1+1/x)^c$ term increases the slope at $x\lesssim1$, and the exponential term causes the cut-off at small $x$, depending on the mass of the halos in question and the parameter $M_c$. We interpret the exponential cut-off as a consequence of the exponential cut-off in the mass function: at $M/x\ll M_c$ this term's contribution is negligible, while at $M/x\gtrsim M_c$ the function decreases exponentially because there is an exponentially small number of halos of mass $M_c$ for halos of mass $M$ to merge with. Indeed, there is a close relationship between $M_c$ and the dark matter halo mass at the knee of the halo mass function $M_*$ (based on EPS; \citealp{MoH_02a}), with $M_c\approx30M_*$. At high redshift, the exponential cut-off affects the function already at $x\lesssim1$, so the parameters $c$, $d$ and $M_c$ become somewhat degenerate. Therefore, we make use of the relationship between $M_c$ and $M_*$ to reduce the freedom of the fitting by setting $\log_{10}(M_c[\Msun])\approx14.8-1.2z$. Table \ref{t:fits} shows the best-fitting numerical values for the parameters of the fitting function, for the curves that appear in Figure \ref{f:rates_x}.

\begin{table}[tbp]
\caption{Fits for the merger rate as a function of mass ratio in Figure \ref{f:rates_x}}
\label{t:fits}
\centering          
\begin{tabular}{c c c c| c c c c c} 
\hline\hline       
Panel & Low mass & High mass & z & A & b & c & d & Mc \\ 
 & $[\Msun]$ & $[\Msun]$ & & & & & & $[10^{10}\Msun]$ \\ 
\hline 
a & $10^{11.25}$ & $10^{11.35}$ & $0.24$ & 0.006 & -0.18 & 0.82 & -0.79 & 32290 \\ 
a & $10^{12.3}$ & $10^{12.7}$ & $0.24$ & 0.011 & -0.32 & 0.57 & -0.89 & 32290 \\ 
a & $10^{13.3}$ & $10^{13.7}$ & $0.24$ & 0.012 & -0.23 & 0.78 & -1.12 & 32290 \\ 
b & $10^{11.25}$ & $10^{11.35}$ & $3.06$ & 0.32 & -0.57 & 0.62 & -0.35 & 13.4 \\ 
b & $10^{12.3}$ & $10^{12.7}$ & $3.06$ & 1.2 & -0.71 & 1.14 & -0.38 & 13.4 \\ 
b & $10^{13}$ & $10^{14}$ & $3.06$ & 2.23 & -0.67 & 0\tablenotemark{(a)} & -0.26 & 13.4 \\ 
c & $10^{12.4}$ & $10^{12.6}$ & $0.24$ & 0.011 & -0.32 & 0.62 & -0.73 & 32290 \\ 
c & $10^{12.4}$ & $10^{12.6}$ & $1.63$ & 0.1 & -0.51 & 0.53 & -0.48 & 698 \\ 
c & $10^{12.4}$ & $10^{12.6}$ & $3.06$ & 1.409 & -0.74 & 1.37 & -0.39 & 13.4 \\ 
\hline                  
\end{tabular}
\tablecomments{(a) For such high mass at high redshift, the exponential cut-off affects the merger rate at $x$ large enough that the upturn provided by the $(1+1/x)^c$ term is absent. The lowest $\chi^2$ is actually achieved in this case with $c<0$, but we do not allow that, hence $c=0$ is forced.}
\vspace{0.5cm}
\end{table}

To investigate the sensitivity of our results to the time resolution of the snapshots in the simulation, we have performed the following test. For any progenitor snapshot $s$, we have extracted the merger rate skipping the two subsequent snapshots, as though the next available snapshot was only $s+3$. At low redshift we find a negligible change in the results. At increasing redshift, $\Delta z$ increases, so when $2$ snapshots are skipped, there occurs a non-negligible effect of averaging the redshift dependence of the merger rate. This is why at $z\approx3$ a decrease of up to $\approx20\%$ can be seen for this "skip $2$" method, a difference that is consistent with being just the result of averaging the redshift dependence. This $\Delta z$-convergence is comparable to the one found by FM08.

As can be seen in Figure \ref{f:rates_x} and Table \ref{t:fits}, the shape of $\frac{1}{N_{\rm prog-halo}}\frac{dN_{\rm merger}}{dtdx}(x,z,M)$ changes with mass and redshift, so that no global values for the fitting parameters $A$, $b$, $c$ \& $d$ can be obtained. This reflects deviation from self similarity, in the sense that halos of different masses and at different redshifts have different weights for merging with halos of different mass ratios. At a given redshift (Figure \ref{f:rate_vs_x_low_z} and Figure \ref{f:rate_vs_x_high_z}), more massive halos have slightly more minor mergers ($x\gtrsim1$, consistently with FM08) while they are being accreted onto larger halos ($x\ll1$) at a much lower rate than less massive halos. Similarly, for a given halo mass (Figure \ref{f:rate_vs_x_const_mass}), at high redshift the minor merger rate (per unit time) is higher, while the rate of being accreted onto bigger halos is lower.

\subsection{The major merger rate per progenitor halo as a function of mass and redshift}
\label{s:rates_M_z}
For a quantitative comparison between similar masses at different redshifts or different masses at fixed redshifts, a specific value or range in $x$ must be chosen. We investigate the dependence on mass and redshift of the possibly most interesting range: $1/3<x<3$, i.e.~the major merger rate:
\begin{eqnarray}
\frac{1}{N_{\rm prog-halo}}\frac{dN_{\rm major-merger}}{dt}(z,M)=\int_{1/3}^{3}\frac{1}{N_{\rm prog-halo}}\frac{dN_{\rm merger}}{dtdx}(x,z,M)dx.
\label{e:major_merger_rate}
\end{eqnarray}
Note that, following the discussion in \S\ref{s:rates}, in cases where integrating over the range $1/3<x<3$ counts the same merger twice, also both halos are counted in $N_{\rm prog-halo}$. In other words, each merger always contributes exactly one count to the merger rate {\it per halo}, so there is no double counting in equation (\ref{e:major_merger_rate}).

\begin{figure*}[tbp]
\centering
\subfigure[]{
          \label{f:rate_vs_M_low_z}
          \includegraphics[]{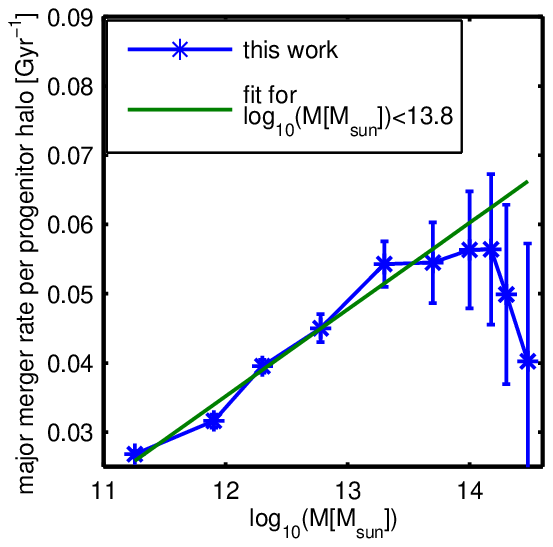}}
\subfigure[]{
          \label{f:rate_vs_M_high_z}
          \includegraphics[]{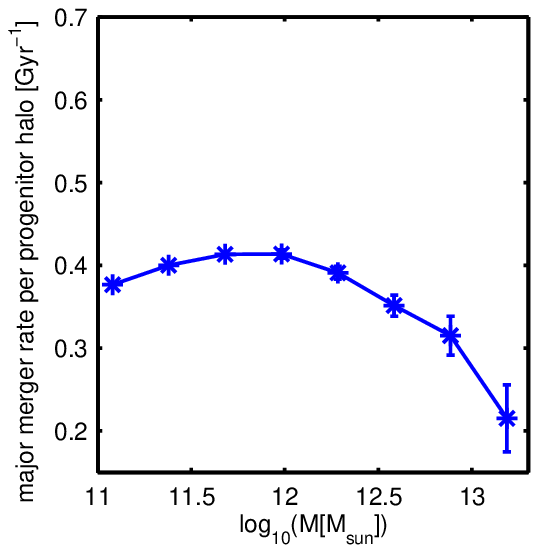}}
\subfigure[]{
          \label{f:rate_vs_z_const_mass}
          \includegraphics[]{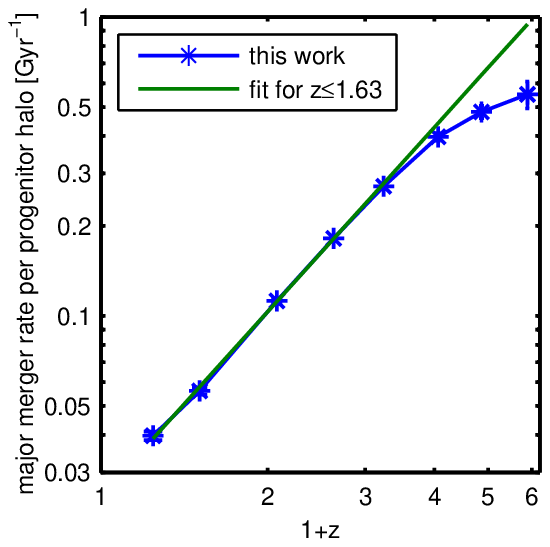}}
\caption{The major merger rate per progenitor halo as a function of mass at $z\approx0.24$ (a) and at $z\approx3$ (b), and as a function of redshift for $10^{12.4}\Msun<M<10^{12.6}\Msun$ (c). The asterisks show our analysis of the simulation, while the solid green curves show the fitting function equation (\ref{e:major_merger_rate_fit}). The fitting functions fit the data well for low enough values of mass and redshift so that equation (\ref{e:critical_mass}) holds. In this regime the major merger rate per unit time increases steeply with increasing redshift and mildly with increasing mass. Above the threshold set by equation (\ref{e:critical_mass}), we find the mass/redshift dependencies to break. Specifically, at high redshift (b) the whole mass range available from the simulation is above the exponential cut-off mass given by equation (\ref{e:critical_mass}), therefore the major merger rate is seen to almost always {\it decrease} with increasing mass. Equation (\ref{e:major_merger_rate_fit}) is not valid in that regime.}
\vspace{0.5cm}
\label{f:MM_rate}
\end{figure*}

\begin{figure*}[tbp]
\centering
\subfigure[]{
          \label{f:rate_vs_M_low_z_comp}
          \includegraphics[]{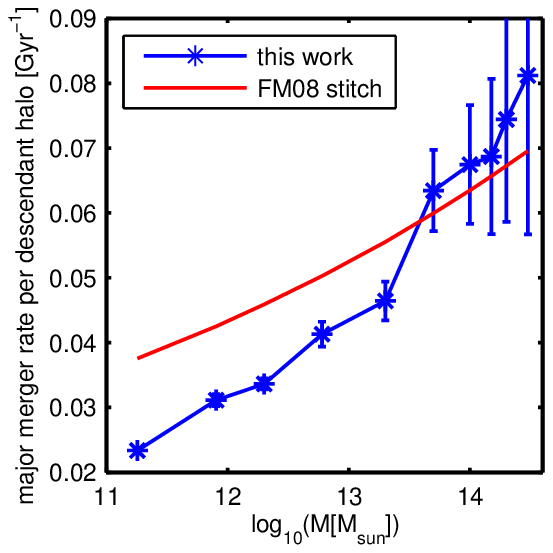}}
\subfigure[]{
          \label{f:rate_vs_M_high_z_comp}
          \includegraphics[]{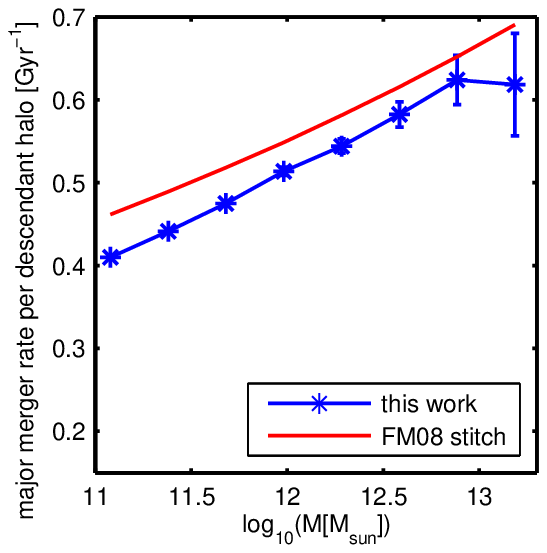}}
\subfigure[]{
          \label{f:rate_vs_z_const_mass_comp}
          \includegraphics[]{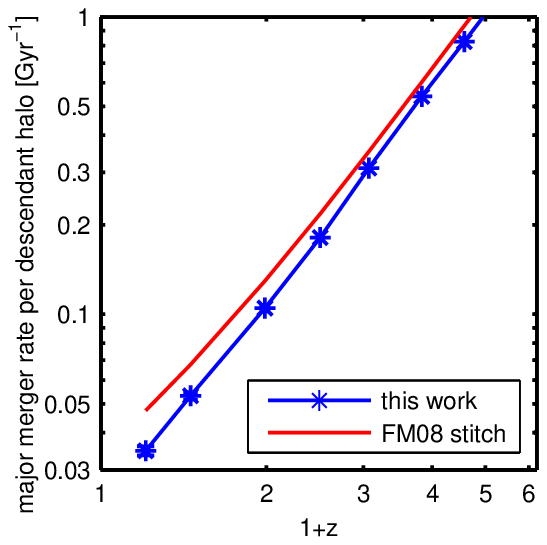}}
\caption{The major merger rate per descendant halo as a function of mass at $z\approx0.24$ (a) and at $z\approx3$ (b), and as a function of redshift for $10^{12.4}\Msun<M<10^{12.6}\Msun$ (c). The asterisks show our analysis of the simulation and the red curves show the merger rate as quantified by FM08's fitting formula (and integrated over the major merger range of mass ratios, between $x=1$ and $x=3$). A comparison to Figure \ref{f:MM_rate} shows that even above the threshold set by equation (\ref{e:critical_mass}) the mass/redshift dependencies of the merger rate per descendant halo do not change, in contrast with the merger rate per progenitor halo. A comparison to FM08's results shows that the major merger rate is lower once our splitting algorithm and our mass definition are used.}
\vspace{0.5cm}
\label{f:MM_rate_comp}
\end{figure*}

Figure \ref{f:rate_vs_M_low_z} shows the dependence of $\frac{1}{N_{\rm prog-halo}}\frac{dN_{\rm major-merger}}{dt}(z,M)$ on mass at $z\approx0.24$. At $M\lesssim10^{13.8}\Msun$ the major merger rate increases with increasing mass, and a relation of the form
\begin{eqnarray}
\frac{1}{N_{\rm prog-halo}}\frac{dN_{\rm major-merger}}{dt}(z,M)\propto\log(M)+\alpha
\label{e:mass_dependence}
\end{eqnarray}
holds, as the solid green line in Figure \ref{f:rate_vs_M_low_z} shows. However, for higher masses the major merger rate {\it decreases} with increasing mass, because the exponential cut-off of $\frac{1}{N_{\rm prog-halo}}\frac{dN_{\rm merger}}{dtdx}(x)$ at small $x$ affects even the range $x>1/3$. At each redshift the critical mass above which equation (\ref{e:mass_dependence}) no longer holds is different. At high redshift there are less massive halos, so the exponential cut-off occurs at $x=1/3$ already for lower masses, compared with low redshift. We find that equation (\ref{e:mass_dependence}) is valid for
\begin{eqnarray}
M\lesssim M_c(z)/5\approx6M_*(z)\approx10^{14.1-1.2z}\Msun.
\label{e:critical_mass}
\end{eqnarray}

Similarly, for a given mass $M$, at low enough redshifts the major merger rate increases with increasing redshift, as shown in Figure \ref{f:rate_vs_z_const_mass}. The relation is of the form $\frac{1}{N_{\rm prog-halo}}\frac{dN_{\rm major-merger}}{dt}(z,M)\propto (1+z)^{\beta}$ with ${\beta}\approx2$. But since $M_*(z)$ decreases with increasing redshift, there is a redshift above which $M>6M_*(z)$. At that redshift the major merger rate begins to be affected by the exponential cut-off, and the power law approximation breaks down. The major merger rate starts {\it decreasing} with increasing redshift only at much higher redshifts (e.g.~while $6M_*(z\approx0.9)=10^{13}\Msun$, the major merger rate of $10^{13}\Msun$ halos starts decreasing with increasing redshift only at $z\gtrsim3$). We note that the regime that is affected by the cut-off exists for the merger rate per progenitor halo that we present in this Section. It does not appear in the merger rate per descendant halo, as will be shown in \S\ref{s:rates_desc} and \S\ref{s:comparison_prog_desc}.

As long as equation (\ref{e:critical_mass}) holds, we find the following fitting function for the major merger rate as a function of mass and redshift:
\begin{eqnarray}
\frac{1}{N_{\rm prog-halo}}\frac{dN_{\rm major-merger}}{dt}(z,M)=R(\mu_{10}+\alpha)(1+z)^{\beta},
\label{e:major_merger_rate_fit}
\end{eqnarray}
with $R=0.0075\Gyr^{-1}$, $\alpha=1\pm0.1$, $\beta=2.1\pm0.1$ and ${\mu_{10}}\equiv\log_{10}(\frac{M}{10^{10}\tiny{\Msun}})$. It is accurate to within $\approx10\%$, an inaccuracy that is included in the errors quoted for the fitting parameters. Because of the inability to properly split halos in the very last snapshots of the simulation at $z\lesssim0.2$ (as explained in \S\ref{s:new_trees}) we do not include this range in our fitting.

An illustration of the regime where equation (\ref{e:critical_mass}) does not hold is shown in Figure \ref{f:rate_vs_x_high_z} and Figure \ref{f:rate_vs_M_high_z}. Since $M_c(z=3)\lesssim10^{11.5}\Msun$, all the mass ranges we can probe at this redshift are in the regime where equation (\ref{e:critical_mass}) doesn't hold. In Figure \ref{f:rate_vs_x_high_z} it is shown that at $z\approx3$ the exponential cut-off occurs at much larger $x$ (compared with Figure \ref{f:rate_vs_x_low_z}), so that also the major and minor merger regimes are affected. This means that the merger rate is {\it lower} for {\it more} massive halos even at $x>1$, as opposed to low redshift, where this is true only for $x\lesssim0.1$. The major merger rate, explicitly as a function of mass, is shown in Figure \ref{f:rate_vs_M_high_z}.

It is interesting to note that the major merger rate per unit redshift,
\begin{eqnarray}
\frac{1}{N_{\rm prog-halo}}\frac{dN_{\rm major-merger}}{dz}(z,M)=\frac{dt}{dz}\frac{1}{N_{\rm prog-halo}}\frac{dN_{\rm major-merger}}{dt}(z,M),
\label{e:major_merger_rate_per_dz}
\end{eqnarray}
is approximately constant with redshift for the range that is not affected by the exponential cut-off, since $\frac{dt}{dz}(z)$ approximately cancels out the redshift dependence in equation (\ref{e:major_merger_rate_fit}). At redshifts higher than the break redshift, the major merger rate per unit redshift decreases steeply with increasing redshift. The bimodal fit:
\begin{eqnarray}
0.38\pm0.02,z\lesssim2.5,\nonumber\\
0.38-0.063(z-2.5),z\gtrsim2.5
\label{e:major_merger_rate_per_dz_fit}
\end{eqnarray}
describes the major merger rate per unit redshift for $10^{12}\Msun$ halos as a function of redshift. It can be applied for different masses using equation (\ref{e:major_merger_rate_fit}) to scale with mass, and equation (\ref{e:critical_mass}) to find the ``break'' redshift. 

If the mass dependence of the major merger rate, which is weak, is neglected, equation (\ref{e:major_merger_rate_fit}) can be integrated between $z_i$ and $z_f<z_i$ to achieve the average number of major mergers throughout the formation history of halos. If a population of halos of mass $M$ is chosen at redshift $z_i$ where $M<6M_*(z_i)$, then for $z<z_i$ the major merger rate per unit redshift is roughly constant with redshift. Therefore the very simple linear relation for the average number of major mergers halos of mass $M$ at redshift $z_i$ will undergo until redshift $z_f$ is
\begin{eqnarray}
\bar{N}_{\rm MM,per-prog}(z_i,z_f,M)\approx0.13\times(\mu_{10}+1)(z_i-z_f),
\label{e:major_merger_number}
\end{eqnarray}
while the number of major mergers individual halos undergo is of course distributed around this average.

\subsection{The merger rate per descendant halo}
\label{s:rates_desc}
We find that the merger rate per descendant halo can be fit much better with a global fitting formula, compared with the rate per progenitor halo. We adopt the fitting form of FM08 (albeit keeping our mass ratio variable $x=1/\xi$), and give different best-fitting parameters that express our different treatment of fragmentations (\S\ref{s:new_trees}) and different mass definition (\S\ref{s:mass_def}):
\begin{eqnarray}
\frac{1}{N_{\rm desc-halo}}\frac{dN_{\rm merger}}{dzdx}(x,z,M)=AM_{12}^\alpha x^b exp((\tilde{x}/x)^\gamma)\frac{d\delta_c}{dz},
\label{e:rate_desc}
\end{eqnarray}
where $M_{12}=M/10^{12}\Msun$ and $\delta_c\approx1.69/D(z)$, which we estimate using the approximation provided by \citet{NeisteinE_08a}.

Our best-fitting parameters are: $A=0.06$, $\alpha=0.12$, $b=-0.2$, $\tilde{x}=2.5$ and $\gamma=0.5$. Note that with the mass ratio definition used by FM08 our parameters correspond, in their notation, to: $A=0.06$, $\alpha=0.12$, $\beta=-b-2=-1.8$, $\tilde{\xi}=1/\tilde{x}=0.4$, $\gamma=0.5$, $\eta=1$ and $\tilde{M}=10^{12}\Msun$. A detailed comparison with the results of FM08 is presented in \S\ref{s:comparison_FM08}.

We find this formula to fit the merger rate per descendant halo per unit redshift with deviations of up to $\approx20\%$ for all the mass range probed by the Millennium Simulation at redshifts $z\lesssim4$. One systematic exception is at $0.5\lesssim z\lesssim1.5$ and $30\lesssim x\lesssim1000$, where the fitting formula tends to overestimate the merger rate by up to $50\%$. At $z\gtrsim4$ the redshift dependence as well as the mass dependence become stronger, and we do not make an attempt to fit that regime.


Also the merger rate per descendant halo can be integrated across cosmic times (neglecting its mass dependence) to obtain the mean number of major mergers that halos of mass $M$ at redshift $z_f$ have undergone since redshift $z_i$. If we approximate $\frac{d\delta_c}{dz}$ as $1.25$ at $z>1$ and $0.8+0.32z$ at $z\leq1$ (an approximation that holds with deviations $<10\%$) then we obtain
\begin{eqnarray}
\bar{N}_{\rm MM,per-desc}(z_i,z_f,M) \approx 0.43M_{12}^\alpha \times \quad \quad \quad \quad \quad \quad \quad \quad \quad \quad \quad \nonumber\\
                                             \left\{ \begin{array}{ll}
                                                        (z_i-z_f) & \hbox{if } 4\gtrsim z_i,z_f\geq1 \\
                                                        z_i-0.65z_f-0.14z_f^2-0.21 & \hbox{if } 4\gtrsim z_i>1,z_f<1 \\
                                                        0.65(z_i-z_f+(z_i^2-z_f^2)/5) & \hbox{if } z_i,z_f\leq1. \\
                                                     \end{array}
                                             \right.
\label{e:major_merger_number_desc}
\end{eqnarray}

\subsection{Comparison between the merger rate per progenitor halo and per descendant halo}
\label{s:comparison_prog_desc}
For mergers of $x\approx1$ at high mass/redshift the different definition of the merger rate, i.e.~per progenitor halo versus per descendant halo (\S\ref{s:rates}), is an important issue. Where equation (\ref{e:critical_mass}) is no longer satisfied, the functional behaviour of the two is different: defining the merger rate per progenitor halo, the exponential cut-off of the merger rate enters into the $x>1$ range, whereas defining it per descendant halo, the function increases all the way to $x=1$, beyond which $x$ is no longer defined. This difference is caused by attributing major mergers (e.g.~between identical halos of mass $M$) to their approximate sum of masses. The same number of mergers is recorded by both methods, but there are far fewer halos with mass $2M$ than halos with mass $M$ in the exponential cut-off regime.

For $x\gg1$, or the minor merger regime, the difference between the two definitions is small. For example, the integrated rate of mergers with $10<x<100$ per progenitor halo is lower by $\approx10\%$ at $z<1$ and $M\approx10^{13}\Msun$ than the same rate per descendant halo.

We further consider the major merger rate, i.e.~the merger rate integrated over $1/3<x<3$ (or $1<x<3$ for the merger rate per descendant halo). Figure \ref{f:MM_rate_comp} shows the merger rate per descendant halo, keeping our halo mass definition and treatment of halo fragmentation ({\it blue, asterisks}). Comparing Figure \ref{f:MM_rate_comp} to Figure \ref{f:MM_rate} illustrates that in the low mass/redshift regime where equation (\ref{e:critical_mass}) holds, the major merger rate is only slightly affected by the merger rate definition. As described in \S\ref{s:rates_M_z}, above the threshold set by equation (\ref{e:critical_mass}), we find the mass/redshift dependencies of the merger rate per progenitor halo to break, a different behaviour than that of the merger rate per descendant halo. At high redshift (Figures \ref{f:rate_vs_M_high_z_comp} and \ref{f:rate_vs_M_high_z}), where all the masses that can be probed using the Millennium Simulation are above the "break mass" (equation (\ref{e:critical_mass})), the different definitions lead to very different results at almost all masses.

Finally, only the merger rate per progenitor halo quantifies the rate of "destruction" events, where halos are accreted onto more massive halos, in an explicit way. This "destruction" merger rate, $x<1$, is not defined by the merger rate per descendant halo alone. To obtain the merger rate with $x<1$ from the merger rate per descendant halo, the halo mass function must be known. In Appendix \ref{s:analytic} we develop analytic formulae that convert merger rates per descendant halo into merger rates per progenitor halo and vice versa, and also relate the merger rate per progenitor halo at $x<1$ to that at $x>1$. The conversion formula from the merger rate per descendant halo to the merger rate per progenitor halo is given by
\begin{eqnarray}
&R_p(z_p,M,x>1)=\nonumber\\
&R_d(z_d,\frac{x+1}{x}M,x)\frac{x+1}{x}N_h(z_d,\frac{x+1}{x}M)/N_h(z_p,M),
\label{e:number_of_mergers_per_prog_halo_b}
\end{eqnarray}
\begin{eqnarray}
&R_p(z_p,M,x<1)=\nonumber\\
&R_d(z_d,(\frac{1}{x}+1)M,\frac{1}{x})(\frac{1}{x}+1)N_h(z_d,(\frac{1}{x}+1)M)x^{-2}/N_h(z_p,M),
\label{e:number_of_mergers_per_prog_halo_small_x_b}
\end{eqnarray}
where $R_p$ is the merger rate per progenitor halo, $R_d$ is the merger rate per descendant halo, $N_h$ is the mass function defined as $N_h(M)\equiv\frac{dn(>M)}{dM}$, and $z_p$ and $z_d$ are the progenitor and descendant redshifts, respectively. We verified numerically that those formulae describe the relations between the different merger rate definitions well.

\section{Comparison to previous work}
\label{s:comparison}
\subsection{Comparison to FM08}
\label{s:comparison_FM08}
In Section \ref{s:new_trees} we have described the differences between our method and FM08's method for building new halo trees and defining the halo mass. We show in this section that those differences have a significant effect on the results. We first discuss the differences between the dependencies on mass ratio, then compare integrated minor and major merger rates, and finally indicate the different time and mass evolutions.

\begin{figure}[tbp]
\centering
\includegraphics[]{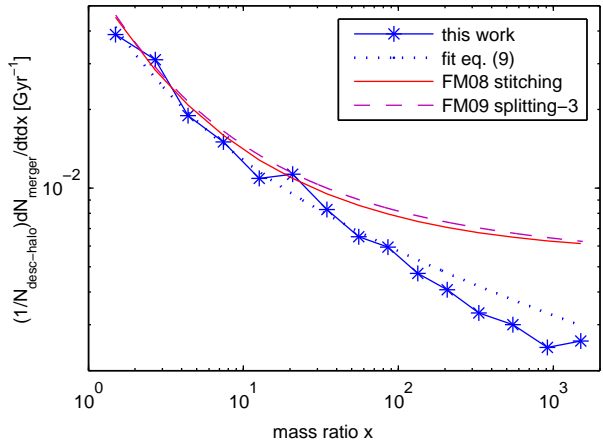}
\caption{The merger rate per descendant halo for $10^{14}\Msun$ halos at $z\approx0.4$, comparing the results of our splitting algorithm to FM08's "stitching" algorithm and FM09's "splitting-3" algorithm (C.-P.~Ma, priv.~comm.), which both give virtually the same results. We find a lower merger rate, starting from $\approx6\%$ at $x\lesssim10$ and increasing steadily to a factor of $2$ at $x\approx1000$. About half of the difference originates in the different halo mass definition (\S\ref{s:mass_def}), and about half from our algorithm that never counts the same merger more than once.}
\vspace{0.5cm}
\label{f:FM08_comp_vs_ratio}
\end{figure}

\begin{figure}[tbp]
\centering
\includegraphics[]{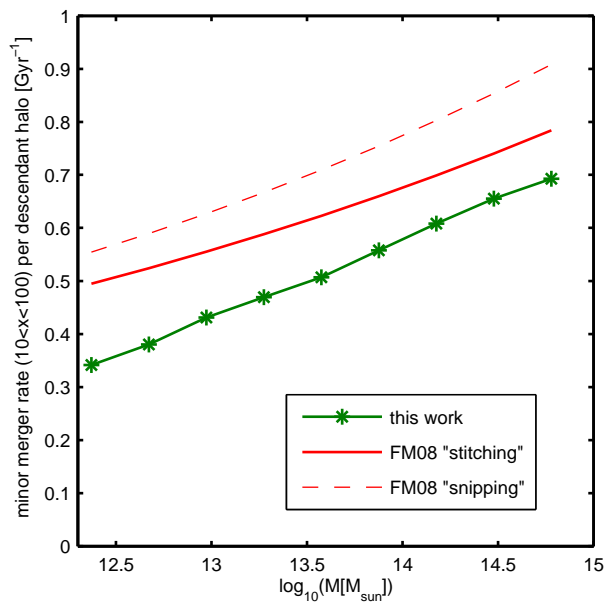}
\caption{The integrated merger rate per descendant halo per unit time of minor mergers between $10:1$ and $100:1$, as a function of mass at $z\approx0.24$. The mass resolution of the simulation allows this rate to be directly measured only for $M\gtrsim10^{12.4}\Msun$. Appropriate treatment of halo fragmentation and an appropriate halo mass definition are significant in this regime. Not giving any special treatment to the fragmentation of halos, thereby counting artificial mergers (FM08's "snipping" method, {\it dashed red}) leads to a significant overestimate of the merger rate. FM08's "stitching" method ({\it solid red}) still overestimates this minor merger rate by a factor $\approx1.3$ compared to our method ({\it green, asterisks}), due to their acceptance as mergers of most temporary links between FOF groups and their definition of a halo mass.}
\vspace{0.5cm}
\label{f:FM08_comp_minors}
\end{figure}

Figure \ref{f:FM08_comp_vs_ratio} shows the merger rate per descendant halo as a function of mass ratio for $10^{14}\Msun$ halos at $z\approx0.4$. For minor mergers FM08\footnote{FM09's "splitting-3" method gives an almost identical merger rate to FM08's fiducial "stitching-3" method, in spite of the differences between the algorithms. An examination of Table \ref{t:counting_mergers} shows that it is probable that those differences compensate each other between the scenarios in the third and fifth columns of that Table.} also find that the merger rate varies as a power law of the mass ratio. However, they find the power law index $b$ to be $\approx0$ ($\beta=-2.01$), i.e.~find more minor mergers. This can be seen at $x\gg1$ in Figure \ref{f:FM08_comp_vs_ratio}. The difference arises because FM08 accept most temporary links between FOF groups as mergers. This leads to an artificial inflation of the number of minor mergers. This was anticipated by FM08 themselves, and is also supported by the finding that subhalos that are ejected out of their host halos are preferentially of low mass \citep{LudlowA_08a}. The second significant parameter that depends on the tree building method is $\tilde{x}$ (see equation (\ref{e:rate_desc})). For the "snipping" method FM08 found that the best-fitting value is $\approx58$, and for the "stitching" method $\approx10$, while we find the best value to be $\tilde{x}=2.5$. This means that the dependence of the merger rate on mass ratio in our method is closer to a pure power-law (since the exponential term affects mostly $x\lesssim\tilde{x}$). The larger the number of false (mostly minor) mergers that are counted, the stronger the exponential term should be, and the steeper the power-law.

The integrated minor merger rate per descendant halo of mergers between $10:1$ and $100:1$ at $z\approx0.24$ as a function of mass is shown in Figure \ref{f:FM08_comp_minors}. It shows that the "snipping" and "stitching" methods ({\it dashed and solid curves}) overestimate the minor merger rate that we find. Roughly half of this difference originates from properly removing all artificial mergers, and equally important is the difference in the mass definition of halos (see \S\ref{s:mass_def}). Note that while Table \ref{t:counting_mergers} shows that $\approx80\%$ of all mergers are treated equally by all methods, a large fraction of all mergers are rather major mergers between rather low mass halos, simply because of the large abundance of low mass halos. Among minor mergers, which involve halos much more massive than the mass resolution limit, many belong to the other $\approx20\%$ that are treated differently by the different methods.

The comparison in the major merger regime can be seen in Figure \ref{f:MM_rate_comp}. The merger rate we find is systematically lower than FM08's also due to our splitting algorithm but, in this regime, mainly due to the different halo mass definition. This is especially true at low mass (Figure \ref{f:rate_vs_M_low_z_comp}) and low redshift (Figure \ref{f:rate_vs_z_const_mass_comp}).

We find the redshift dependence of the merger rate to be proportional to $\frac{d\delta_c}{dz}$, while FM08 find a slightly weaker redshift evolution at low redshift by introducing $(\frac{d\delta_c}{dz})^{\eta}$ with $\eta\approx0.3$. The consequence of $\eta=1$ is discussed further in \S\ref{s:comparison_EPS}. We also find a slightly stronger mass dependence, $\alpha=0.12$ rather than $\alpha\approx0.08$.

\subsection{Comparison to the EPS model}
\label{s:comparison_EPS}
\begin{figure*}[tbp]
\centering
\subfigure[]{
          \label{f:comparison_EPS_vs_x}
          \includegraphics[]{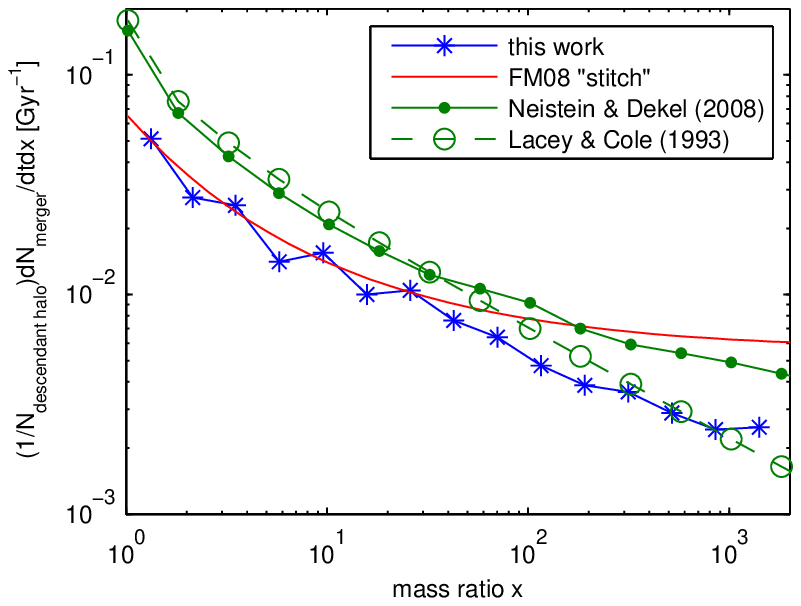}}
\subfigure[]{
          \label{f:comparison_EPS_vs_z}
          \includegraphics[]{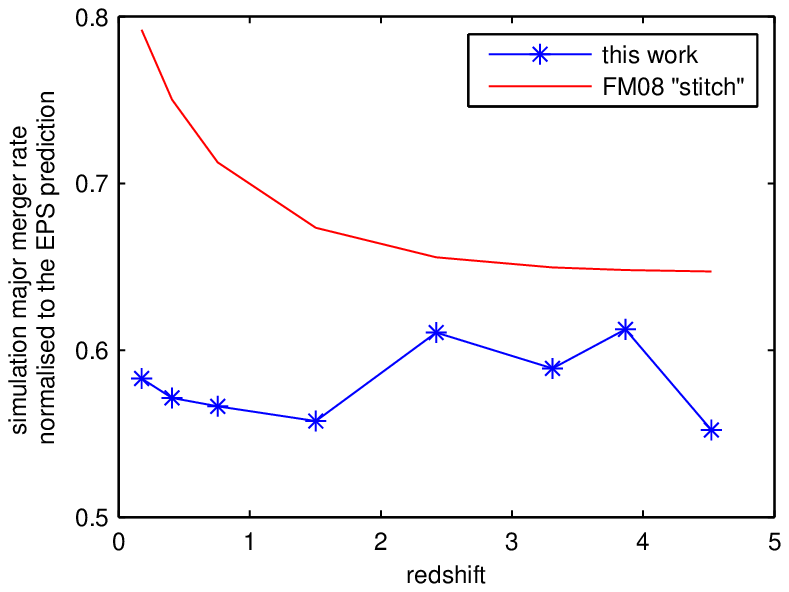}}
\caption{A comparison between different results for the merger rate per descendant halo per unit time. Panel (a) illustrates the merger rate for $M\approx10^{14}\Msun$ at $z\approx0.4$. Four different methods are shown: using the splitting algorithm and the mass definition presented in \S\ref{s:find_merger} ({\it blue, asterisks}), FM08's fit to their "stitching" method ({\it dashed red}), the \citet{NeisteinE_08b} EPS method ({\it green, filled circles}) and the \citet{LaceyC_93a} EPS method ({\it dashed green, open circles}). At $x\lesssim30$, the two EPS methods are similar and exceed the two methods based on the Millennium Simulation by $\approx70\%$. As $x$ increases towards more minor mergers, FM08's merger rate is the shallowest, finding the highest number of minor mergers, while the EPS method of \citet{LaceyC_93a} is the steepest, finding the lowest number of minor mergers. On the other hand, the slopes found by our analysis and by the EPS method of \citet{NeisteinE_08b} are similar, so that a roughly constant ratio is kept between the inferred merger rates (\S\ref{s:comparison_EPS}). Panel (b) compares the integrated major merger rate ($1\leq x\leq3$) of $M\approx10^{12.5}\Msun$ halos, as a function of redshift, that results from our method ({\it blue, asterisks}), FM08's "stitching" method ({\it dashed red}) and the \citet{NeisteinE_08b} EPS method, with which both other methods are normalised. The redshift dependence resulting from our analysis of the simulation agrees well with that of the EPS prediction (yet, again, with an offset in the normalisation), while FM08 find more mergers at low redshift because their method is more sensitive to artificial mergers, which are more common at low redshift.}
\vspace{0.5cm}
\label{f:comparison_EPS}
\end{figure*}

Figure \ref{f:comparison_EPS} compares different results for the merger rate per descendant halo of $10^{14}\Msun$ halos at $z\approx0.4$. The merger rate per descendant halo is the natural quantity that is obtained from the EPS model, and does not require explicitly the mass function of halos.

FM08 found that the EPS model predicts more major mergers and (especially) fewer minor mergers compared with the Millennium Simulation. The slope of the merger rate at $x\gg1$ is determined by the parameter $b$ in equation (\ref{e:rate_desc}). The larger it is, the more minor mergers dominate the number of mergers. Indeed, FM08's "snipping" method, which doesn't reject any artificial mergers, finds $b=0.17$, while their fiducial "stitching" method (red curve in Figure \ref{f:comparison_EPS_vs_x}) finds $b=0.01$. The value we find, by rejecting {\it all} artificial mergers (blue curve with asterisks in Figure \ref{f:comparison_EPS_vs_x}), is $b=-0.2$, while the \citet{LaceyC_93a} EPS prediction (green curve with open circles in Figure \ref{f:comparison_EPS_vs_x}) is $b\approx-0.5$. Therefore, we find, like FM08, that the EPS model overpredicts the major merger rate, but we find a better agreement with EPS in the minor merger regime, because we don't accept artificial minor mergers.

Recently \citet{NeisteinE_08b} have constructed a new method for predicting merger rates from the EPS model, by avoiding the assumption of binary mergers, which leads to inconsistency within EPS. They find, in agreement with the simulation, that the \citet{LaceyC_93a} merger rates are too low in the minor merger regime. Moreover, we find that the \citet{NeisteinE_08b} merger rate (green curve with filled circles in Figure \ref{f:comparison_EPS_vs_x}) has at $x\gg1$ a slope of $b=-0.19\pm0.02$, i.e.~its shape agrees remarkably well with what we find in the Millennium Simulation. The \citet{NeisteinE_08b} merger rate is higher than the merger rate we find in the Millennium Simulation by a factor of $\approx1.5\pm0.3$, for all mass ratios, halo masses and redshifts, with almost no systematic dependence on any of those parameters.

This result is further demonstrated in Figure \ref{f:comparison_EPS_vs_z}. The \citet{NeisteinE_08b} merger rate is constant with respect to the natural dimensionless EPS time variable $\delta_c$, which is exactly the same dependence we find in the simulation. Figure \ref{f:comparison_EPS_vs_z} compares the major merger rate per descendant halo of $M\approx10^{12.5}\Msun$ halos between FM08's "stitching" method, the method presented here, and \citet{NeisteinE_08b}. The redshift dependence of FM08's method is significantly different at low redshift (originating numerically from their best-fitting value of $\eta\approx0.3$), because artificial mergers appear preferentially at low redshift. The redshift dependence of the merger rate in the Millennium Simulation based on our method matches well that of EPS (see also \citet{NeisteinE_08a}), but the normalisation, as already mentioned, has an offset.

\subsection{The role of the halo mass definition}
\label{s:mass_def}
There is a substantial uncertainty as to how to determine the boundaries of halos in N-body simulations. Sometimes the mass of halos is taken as the mass inside a sphere, within which the density equals the expected density of virialised groups in the spherical collapse model. This class of definitions is not well-suited for our purposes, because they are less reliable for halos undergoing mergers \citep{WhiteM_01a,LukicZ_08a}. Particles grouped together by the FOF algorithm are considered to correspond to virialised dark matter halos, because their average density approximately equals the expected density of virialised groups in the spherical collapse model, yet the halos can have any shape and are not assumed to be spherical. On the other hand, undesired effects like particle bridges between halos (as discussed in \S\ref{s:new_trees}) and spurious linkage of particles to groups are inherent to the algorithm. An advantage of SUBFIND over FOF is that it subjects the (geometrically-identified) groups of particles to a dynamical test. Only particles that are found to be gravitationally bound to their subhalo are included as part of their subhalo's mass. Therefore, two reasonable definitions we consider for the mass of a halo are: (1) the total mass of all the particles associated with the halo by the FOF algorithm (as in e.g.~FM08), and (2) the total mass of all the particles that belong to the halo and are also gravitationally bound to any of its subhalos.

As described in \S\ref{s:new_trees}, for the population of halos as a whole, the difference between those two definitions is not large. For example, the average fraction of unbound particles is $\approx3\%$. Nevertheless, there is a distinct population of halos for which the typical fraction is much larger. These are halos that are about to undergo a significant merger, i.e.~undergo a major merger or be accreted onto a more massive halo. First we describe how this affects the merger rate, and then interpret this phenomenon and justify our choice of not including the unbound particles when computing the merger rate.

\begin{figure}[tbp]
\centering
\includegraphics[]{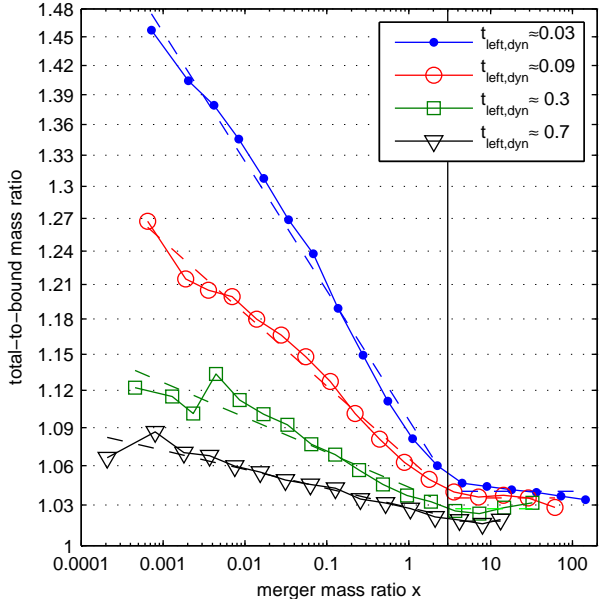}
\caption{The ratio of the total FOF mass of halos to their total SUBFIND bound mass, as a function of the mass ratio of the merger they are about to undergo. The more massive the other halo (the smaller $x$), the larger the total-over-bound mass ratio enhancement, but minor mergers ($x>3$, to the right of the horizontal line) have almost no effect. The more imminent the merger, normalised to the dynamical time, the larger its influence on the halo's total-over-bound mass ratio. These trends are shown by $4$ different curves, each representing halos that are about to undergo such a merger in a fixed "normalised time" into their future, as indicated by the legend. All halos shown here are at the same snapshot, $z\approx3$, but the fitting formulas equations (\ref{e:ratio_vs_ratio}) and (\ref{e:ratio_vs_ratio2}), which are shown by the dashed curves, hold for all redshifts. Since $H(z\approx3)\approx0.3\Gyr^{-1}$, the curves in this figure correspond to halos that are about to undergo a merger in $0.1,0.3,1,2.4\Gyr$ into their future, which corresponds to $1,2,5,10$ simulation snapshots, from top to bottom respectively.}
\vspace{0.5cm}
\label{f:ratio_vs_ratio}
\end{figure}

Halos at snapshot $s$ that are about to undergo a merger have on average higher ratios of total FOF mass over bound SUBFIND mass. We find that the total-over-bound mass ratio of a halo correlates with two quantities related to a halo's next merger: 1) the time before the next merger begins, relatively to the dynamical time $\propto H(z)^{-1}$, and 2) the mass ratio of that merger. Specifically, as a halo approaches a merger with another halo of comparable or larger mass, its total-over-bound mass ratio increases with time, an effect that is stronger as the mass of the other halo is larger. This phenomenon can be well quantified for $x<=3$ by
\begin{eqnarray}
\frac{M_{FOF}}{M_{bound}}(x<=3)=1.02\times t_{left,dyn}^{-0.018}\times x^{-0.005\times t_{left,dyn}^{-0.55}},
\label{e:ratio_vs_ratio}
\end{eqnarray}
where $t_{left,dyn}\equiv t_{left}H(z)$, and $t_{left}$ is the time left before the merger starts (which is defined for this purpose as the middle of the time interval between the adjacent snapshots of the progenitors and the descendant). $\frac{M_{FOF}}{M_{bound}}$ denotes the geometrical mean over all halos, a quantity that is used because the scatter is large. Halos that are approaching a minor merger ($x>3$) show almost no enhancement of the total-over-bound mass ratio, therefore we fit it as
\begin{eqnarray}
\frac{M_{FOF}}{M_{bound}}(x>3)=1.02\times t_{left,dyn}^{-0.007}.
\label{e:ratio_vs_ratio2}
\end{eqnarray}
Figure \ref{f:ratio_vs_ratio} demonstrates this phenomenon and its description by equations (\ref{e:ratio_vs_ratio}) and (\ref{e:ratio_vs_ratio2}) for halos at $\approx3$.

\begin{figure}[tbp]
\centering
\includegraphics[]{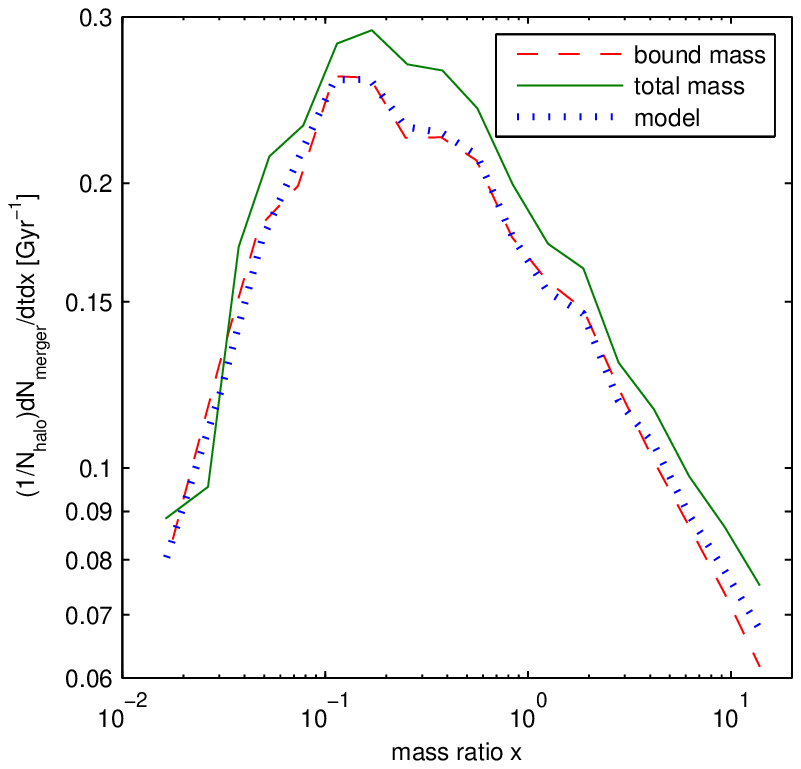}
\caption{The merger rate per progenitor halo for halos of $M\approx10^{12}\Msun$ at $z\approx3$, extracted using the total FOF group mass as the halo mass ({\it solid green}) and using only the bound mass as the halo mass ({\it dashed red}). When the halo mass is defined as the total FOF mass, including gravitationally unbound particles, the inferred merger rate is higher by up to $50\%$ (in a similar way for all masses at all redshifts). This is caused by the fact that halos that are about to merge with more massive halos have unusually high total-to-bound mass ratios. To check whether this is the only source of difference, we used the merger trees in which the mass is defined as the total mass, and changed them according to equations (\ref{e:ratio_vs_ratio}) and (\ref{e:ratio_vs_ratio2}), computing $t_{left,dyn}$ for $z\approx3$ and a $1$ snapshot difference. Specifically, each halo's mass was reduced by $4.55\%$, except for the mass of halos about to undergo a merger of mass ratio $x<=3$, which was reduced by a factor of $1.087x^{-0.0348}$. Although this model doesn't include the scatter in the total-over-bound mass ratio, the result ({\it dotted blue}) is very similar to the results achieved directly with each halo's bound mass. The integral of the merger rates with both mass definitions between $x=0$ and $x=M/M_{min}$ is by construction equal. Nevertheless, we restrict our results to $x\leq M/(2\times M_{min})$, so the sharp, artificial, drop of the solid green curve at $x>M/(2\times M_{min})$ is not seen.}
\vspace{0.5cm}
\label{f:rate_compare_total_and_bound_mass}
\end{figure}

Since the more massive halo of the merging pair (or group) retains the typical value of $\approx1.03$, and the less massive halos have untypically large total-over-bound mass ratios, the mass ratios of mergers shift towards smaller values (mergers become more equal-mass) when the total mass is taken into account, compared with the choice of the bound mass as the halo mass. The effect on the merger rate is pronounced. When the halo mass is taken as the total mass, there is a deficiency of very high mass ratio mergers near the resolution limit, but since we restrict our analysis to mergers with halos of $M/x>2\times M_{min}$, this is not seen. What is seen is an enhancement of the merger rate for every $x<M/(2\times M_{min})$. The merger rate using the total FOF mass is larger by typically $20\%$ and up to $\approx50\%$ than our fiducial method. Figure \ref{f:rate_compare_total_and_bound_mass} demonstrates this difference. Furthermore, Figure \ref{f:rate_compare_total_and_bound_mass} shows that using the mean relations equations (\ref{e:ratio_vs_ratio}) and (\ref{e:ratio_vs_ratio2}) to interchange between the two halo mass definitions allows reproducing the different resulting merger rates.

\begin{figure}[tbp]
\centering
\includegraphics[]{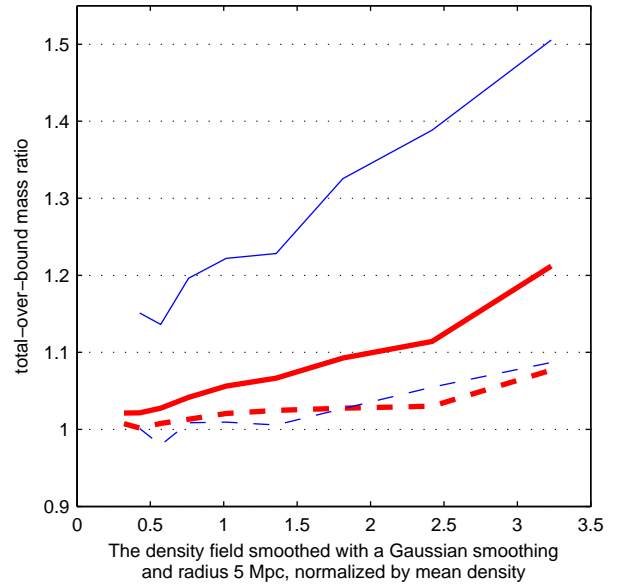}
\caption{The total-over-bound mass ratio of FOF groups as a function of their environment and merging state. Halos residing in denser environments have on average higher total-over-bound mass ratios ({\it thick red}). The same is true for a sub-population of halos that are just about to merge onto a more massive halo ({\it thin blue}). Also, for a given overdensity, halos that are just about to merge onto a more massive halo have significantly higher total-over-bound mass ratios, in accordance with Figure \ref{f:ratio_vs_ratio}. Yet, once each halo's total-over-bound mass ratio is normalised by the value expected for it by equations (\ref{e:ratio_vs_ratio}) and (\ref{e:ratio_vs_ratio2}), the correlation with the environment almost disappears ({\it dashed curves}). This figure shows halos at $z\approx1$, but we find the same trends at any redshift, and for any available choice of a smoothing length: $1.25,2.5,5,10\Mpc$.}
\vspace{0.5cm}
\label{f:mass_vs_density}
\end{figure}

The total-over-bound mass ratio of halos correlates also with the environment in which they reside. Figure \ref{f:mass_vs_density} shows that halos in denser environments have higher total-over-bound mass ratios ({\it solid curves}). This trend holds also among halos that are about to undergo a $x<1$ merger in the following snapshot ({\it thin blue}). Yet, once each halo's total-over-bound mass ratio is normalised by the value expected for it by equations (\ref{e:ratio_vs_ratio}) and (\ref{e:ratio_vs_ratio2}), the correlation with the environment almost disappears ({\it dashed curves}). The difference, at a given overdensity, between all halos and halos about to undergo a $x<1$ merger shows that it is not possible to use the correlation with the environmental overdensity to disentangle the dependence shown in Figure \ref{f:ratio_vs_ratio}. This means that the environmental dependence is probably just a second-order correlation, originating from the correlation shown in Figure \ref{f:ratio_vs_ratio} in combination with the correlation between mergers and environment (cf.~FM09).

\begin{figure}[tbp]
\centering
\includegraphics[]{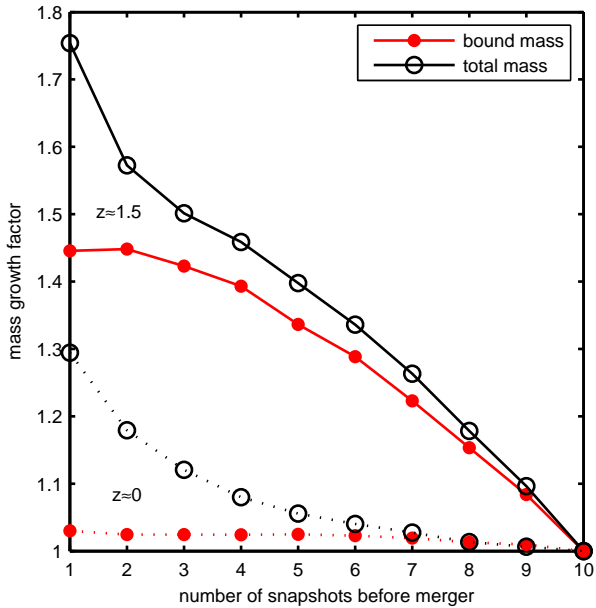}
\caption{Mean mass growth histories for halos that are about to undergo a merger of $0.0005<x<0.01$ at $z\approx0$ ({\it dotted}) or $z\approx1.5$ ({\it solid}), for $10$ subsequent snapshots before the merger. Both the total FOF group mass ({\it empty circles}) and the bound mass ({\it filled circles}) are shown. Each curve is normalised to $1$ at the time of $10$ snapshots before the merger. It is seen that the total mass grows faster than the bound mass, which results in increased total-over-bound mass ratios as these halos approach a merger with a more massive halo. For halos not about to undergo a significant merger, the total and bound mass grow at the same rate, thereby conserving the mean ratio of $\approx1.03$.}
\vspace{0.5cm}
\label{f:mass_histories}
\end{figure}

The strong correlation of the total-over-bound mass ratio of halos with their proximity to their next merger and its mass ratio, and the fact that there is almost no additional dependence on environment, suggest that the mergers themselves are responsible for the change in the total-over-bound mass ratio. The question to be asked is whether this is a gravitational effect, which changes the bound mass even before the FOF groups merge, or a numerical effect, which changes the FOF group mass even before the gravitational interaction is significant. To give a fully satisfactory answer to that question, a detailed dynamical analysis is needed, which cannot be obtained with just the merger trees, and is outside the scope of this paper. Nevertheless, Figure \ref{f:mass_histories} shows that it is not the bound mass that decreases before the merger, but the total mass that increases more rapidly, that causes the total-over-bound mass ratio to increase. We postulate that a significant amount of particles are added to the outskirts of the FOF group during the few snapshots before the merger, but are not found to be bound to it. Therefore, we favour the interpretation of the bound mass as the "true" mass of the halo.

\section{Comparison to observations}
\label{s:merger_fractions}
The two different definitions of the merger rate, per progenitor halo and per descendant halo, have different physical meanings, and they correspond to the two different observational approaches towards measuring the galaxy merger rate. When the merger rate is measured via pair counting, the mass (or luminosity) of each galaxy in the pair is measured separately, and the merger fraction/rate is attributed to the measured progenitor galaxy mass. The merger rate in this case is related to the time scale for galaxies of a given mass/luminosity to encounter other galaxies of a given mass/luminosity, analogously to the merger rate per progenitor halo. In the case of morphological/kinematical identification of disturbed galaxies that show signs of mergers, it is difficult to infer the mass of each of the original components that have merged, and the merger is then attributed to the total mass of the system, i.e.~the descendant mass. In that case, the merger rate is related to the time scale on which a population of galaxies is created by mergers. While such observations still suffer from large uncertainties, it is important to realise that the merger rates that are inferred by both methods are not the same quantity. Those two quantities are related to each other by our conversion formula, in a fully analogous way to the two definitions of the merger rate of dark matter halos.

Some evidence for the expected difference between the two methods can be found in the literature. For example, \citet{MallerA_06a} measure the galaxy major merger rate in a cosmological hydrodynamic simulation, and define it as being per descendant galaxy. They find a steep monotonic relation between increasing mass and increasing merger rate, similar to the trend of the halo merger rate per descendant. Also \citet{ConseliceC_08a}, measuring the merger fraction using a morphological investigation, find that the merger fraction (at $z\gtrsim1.5$) increases strongly with increasing mass. On the other hand, \citet{PattonD_08a} find that the major merger rate of galaxies, obtained via observed pair counting, peaks with respect to luminosity, therefore also with respect to mass. This different dependence, which at first might seem to be at odds with the findings of \citet{MallerA_06a} and \citet{ConseliceC_08a}, is actually qualitatively expected once it is taken into account that \citet{PattonD_08a} measure (implicitly) the merger rate per {\it progenitor} galaxy. Therefore, when comparing either observed or simulated galaxy merger rates, the nature of the observations and the definition used to analyse the simulations must be taken into account. Similarly, \citet{ConseliceC_08a} find that the redshift dependence of the merger fraction becomes stronger for higher mass galaxies, while \citet{DeRavelL_08a}, measuring the pair fraction of galaxies, find exactly the opposite trend, i.e.~that higher mass galaxies have a shalower redshift evolution. This is again in qualitative agreement with our expectations based on the difference between merger rates per progenitor and per descendant halo/galaxy.

\begin{figure}[tbp]
\centering
\includegraphics[]{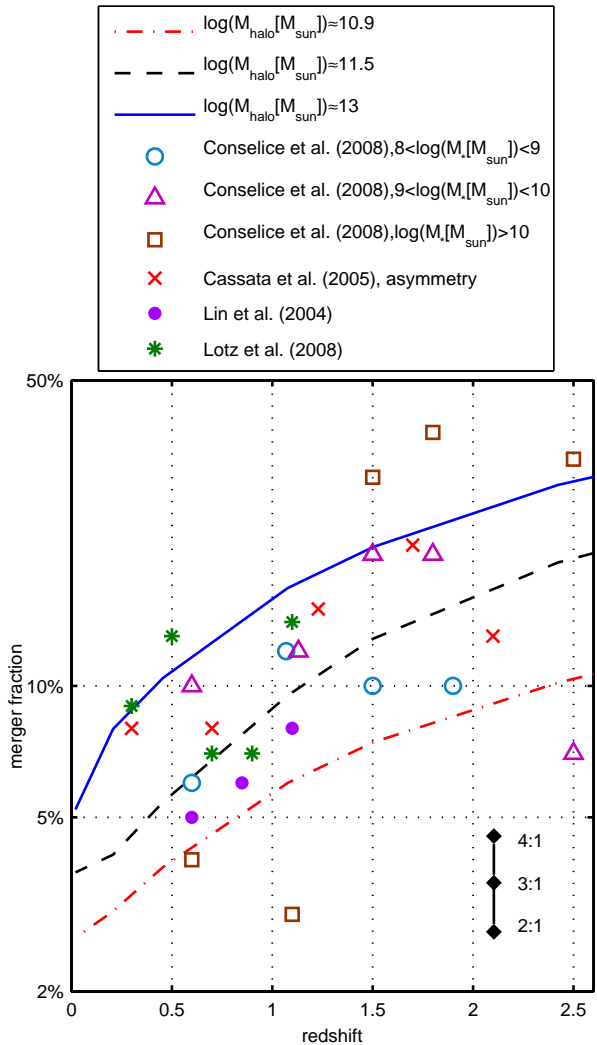}
\caption{The merger fraction as a function of redshift. Values for dark matter halos undergoing major mergers (with a mass ratio threshold of $3$) in the Millennium Simulation are shown for 3 different halo masses, starting from the lowest halo mass we can reliably probe, that is $\approx10^{10.9}\Msun$. The sensitivity to the chosen mass ratio threshold is indicated in the lower right corner. A compilation of observational values that were obtained in different methods and with different selections nevertheless shows a rough agreement, also with the results from the simulation. The error bars of the observations are largely comparable or smaller than the spread, and for the simulation (Poisson errors) they are very small. They have been suppressed to allow better readability.}
\vspace{0.5cm}
\label{f:merger_fractions}
\end{figure}

To quantitatively compare observations with the frequency of galaxy mergers predicted by simulations, a treatment of baryonic physics must be included. Such a comparison is outside the scope of this paper, but can be found in e.g.~\citet{BertoneS_09a} and P.~F.~Hopkins et al.~(2009, in preparation). However, we make here a comparison between the halo merger fraction (per descendant halo) in the Millennium Simulation and galaxy merger fractions (per descendant galaxy) from observations. In Figure \ref{f:merger_fractions} we show a compilation of observed galaxy merger fractions as a function of redshift. Two features are apparent: the merger fraction increases with redshift, and the scatter at any given redshift is roughly a factor of $5$. If we distinguish between fractions according to a rough luminosity/mass criteria, the large scatter remains. This is because the scatter is caused by numerous factors, e.g.~selection of different populations and different wavebands, different techniques for identifying mergers, as well as cosmic variance.

To extract merger fractions from the simulation, a time scale must be associated with mergers (see also \citet{GenelS_08a}). We use the dynamical friction time scale for dark matter halo mergers found by \citet{Boylan-KolchinM_07a} based on merger simulations. We average the orbital parameters and thereby obtain $T_{\rm merger}=0.7\frac{r^{1.3}}{ln(1+r)}\frac{H(z=0)}{H(z)}\Gyr$, where $r$ is the mass ratio and $H(z)$ is the Hubble constant at redshift $z$. Whenever a major merger occurs between snapshots $s$ and $s+1$ (\S\ref{s:rates}), we tag the most massive progenitor at snapshot $s$, and its descendants in following snapshots, as "undergoing a major merger" for a time $T_{merger}$ after snapshot $s$. This allows us to determine, for any snapshot and halo mass range, the fraction of halos that are instantaneously undergoing major mergers, i.e.~the major merger fraction. In Figure \ref{f:merger_fractions} we plot the resulting major merger fraction as a function of redshift for three halo masses.

We note that by tagging the progenitor halo only at snapshot $s$ and the descendant halos in the following snapshots for a duration $T_{merger}$, we derive a merger fraction that is mostly "per descendant halo". We do that for two reasons. First, while a halo that is a descendant of a merger is a halo that hosts galaxies that are about to merge, it is probably observationally less relevant when two halos are still approaching their merger. Second, it is unclear what time scale should be used for an imminent halo merger. The dynamical friction time scale is relevant only after the halos have started coalescing \citep{Boylan-KolchinM_07a}. In other words, there is no proper definition for a "halo pair". We also note that the exact algorithm used for building the merger tree and the halo mass definition become relatively unimportant in this comparison due to the large uncertainties. \citet{StewartK_08b}, who performed a similar analysis, have also shown that it is possible to find a good rough agreement with observations while having much freedom in parameters like mass ratio and time scale.

The theoretical expectation for the dark matter halo merger fraction brackets the observations reasonably well. This suggests that, at least within the current uncertainties, the galaxy merger fraction roughly follows the dark matter halo merger fraction. This conclusion may seem to be different from the findings of \citet{GuoQ_07a}, namely that the role of major mergers in galaxy growth is different from that in halo growth, but the two conclusions are actually consistent with each other. \citet{GuoQ_07a} found that the specific formation rate of halos through major mergers increases steeply with redshift but depends weakly on mass, and vice versa for galaxies. Nevertheless, the specific formation rate they introduced is a different quantity from the merger fraction we show in Figure \ref{f:merger_fractions}. \citet{GuoQ_07a} defined the specific formation rate such that it equals the merger rate per descendant halo as defined in this work, times the Hubble time, with no dependence on the duration of mergers. Thus, the difference they found between halos and galaxies originates solely from the difference of the merger {\it rate} between halos in the dark matter simulation and the galaxies in the semi-analytical model. \citet{GuoQ_07a} found that the galaxy specific formation rate via major mergers is roughly constant with redshift, therefore by dividing it by the Hubble time, we learn that the galaxy major merger rate scales roughly as $(1+z)^{1.3}$. In comparison, they find that the halo specific formation rate via major mergers scales as $(1+z)^{1}$, which after being devided by the Hubble time means that the halo major merger rate scales as $(1+z)^{2.3}$. This is consistent with our finding that the halo major merger rate scales roughly as $(1+z)^{2}$. However, the merger {\it fraction} is proportional also to the merger time scale. The time scale of halo mergers $T_{merger}$ scales roughly as $(1+z)^{-1.3}$ via its dependence on the Hubble time, while the observed time scale for galaxy mergers is probably approximately constant with redshift (e.g.~\citealp{ConseliceC_06a,KitzbichlerM_08a}). Combining these dependencies, one finds that both the halo merger fraction and the galaxy merger fraction scale roughly as $(1+z)^{1}$, which is indeed what is seen on average in Figure \ref{f:merger_fractions}.

\section{Summary and discussion}
\label{s:summary}
We have used the Millennium Simulation to extract merger rates of dark matter halos. Our method differs from previous work in three main aspects.

First, we reject any merger between FOF groups whose descendant subhalos, at any future time, do not belong exclusively to the same FOF group. This is done by keeping such FOF groups distinct until they (if at all) irrevocably merge. Rejecting only a fraction of such events, as in previous work, leads to double counting of mergers, and to false counting of fly-by events as mergers. Therefore, our method results in a lower merger rate, especially in the minor merger regime. \citet{LudlowA_08a} find that ejections of low mass subhalos out of their host halos typically occur in a configuration where a bound group of subhalos (which was created via past mergers) is accreted onto a large halo, and its low mass members are propelled onto high energy orbits by the multiple-body interaction. In the context of merger counting, it is not clear whether the merger of the low mass halo with the group should be counted as a merger, or merely regarded as an interaction that was interrupted by the accretion of the group onto the large halo. Our method assumes the latter for reasons described in \S\ref{s:new_trees}. Therefore, in a more conservative approach, our results can be regarded to as a strong lower limit for the merger rate.

Second, we define each halo's mass as the mass of all the particles gravitationally bound to it, rather than of all the particles constituting the FOF group. This definition reduces the inferred merger rate by typically $20\%$. The motivation for this definition is our finding that the total (bound \& unbound) mass of FOF groups artificially increases by up to $50\%$ as they approach more massive FOF groups on their way to merge with them. This effect distorts the appropriate mass ratios of mergers, thereby changing the merger rate. A detailed discussion of this issue appears in \S\ref{s:mass_def}.

These two improvements make our inferred merger rate more consistent with the new predictions of \citet{NeisteinE_08b} based on the EPS model, in the sense that the functional dependencies of the merger rate on mass ratio, halo mass and redshift are very similar. There is a constant factor of $\approx1.5$ between the EPS merger rate and the Millennium Simulation merger rate. In \S\ref{s:rates_desc} we provide a simple global fitting formula (equation (\ref{e:rate_desc})) for the merger rate per descendant halo that holds for $z\lesssim4$ and all masses probed by the Millennium Simulation.

Third, we also extract the merger rate per {\it progenitor} halo. This allows us to find a merger rate for the full range of mass ratios. For halos of any mass $M$, we can measure the rate at which they undergo minor mergers, major mergers and the rate at which they are being accreted as satellite halos onto more massive halos. This has significant implications for the redshift and mass dependencies of the major merger rate. We find the merger rate, in the regime where both halo masses are smaller than the knee of the mass function, to increase steeply with redshift and only slightly with mass, in a similar way to what was found in previous work. However, at high enough mass or redshift, the number of halos decreases exponentially and therefore the mass dependence of the major merger rate changes sign and starts decreasing with increasing mass. Also, the redshift dependence significantly weakens. In \S\ref{s:comparison_prog_desc} we provide an analytic expression for converting merger rates per descendant halo to merger rates per progenitor halo, which can be used for any theoretical merger rate.

The two different definitions of the merger rate, per progenitor halo and per descendant halo, have different physical meanings. They also correspond to the two different observational approaches towards measuring the galaxy merger rate, namely pair counting and morphological/kinematical identification. In \S\ref{s:merger_fractions} we discuss the importance of this distinction and its relevance to observations. Finally, we find that observed galaxy merger fractions are consistent with the halo merger fraction in the Millennium Simulation within the large observational uncertainties and uncertainties in parameters (like mass ratio or time scale) needed to compare the two. More refined comparisons still await significant improvements both in observational techniques and consistency and in the theoretical treatment of baryonic physics.

\acknowledgements
We thank Eyal Neistein for kindly providing his code for computing EPS-based merger rates, and for very useful discussions. We also thank Mike Boylan-Kolchin for fruitful discussions. The Millennium Simulation databases used in this paper and the web application providing online access to them were constructed as part of the activities of the German Astrophysical Virtual Observatory. We are grateful to Gerard Lemson who devotedly helped us to use the public databases. We thank the referee, Onsi Fakhouri, for an extensive and insightful report that contributed much to the quality of this manuscript. SG acknowledges the PhD fellowship of the International Max Planck Research School in Astrophysics, and the support received from a Marie Curie Host Fellowship for Early Stage Research Training. We thank the DFG for support via German-Israeli Project Cooperation grant STE1869/1-1.GE625/15-1.

\appendix
\section{An analytic conversion formula from merger rates per descendant halo to merger rates per progenitor halo}
\label{s:analytic}
The merger rate per halo equals, in general, the number of merger events divided by the number of halos undergoing those mergers. Let us denote by $r_d(z_d,M,x)dMdx$ the number of mergers\footnote{Per unit time or redshift and possibly per unit volume. This has no significance for this derivation, as long as it is kept consistent throughout.} that occur between $z_p$ and $z_d$, whose descendant mass is $M\pm dM/2$ and ratio is $x\pm dx/2$. Similarly the number of mergers whose more massive progenitor mass is $M\pm dM/2$ and ratio is $x\pm dx/2$ are denoted as $r_p(z_p,M,x)dMdx$, and accordingly for the less massive progenitor: $r_{p2}(z_p,M,x)dMdx$. We will define the halo mass function is the following way: if $n(z,>M)$ is the number of halos more massive than $M$ at redshift $z$, then the number of halos in the interval $M\pm dM/2$ is $\frac{dn(z,>M)}{dM}dM\equiv N_h(z,M)dM$. For simplicity, in the following derivation we assume the binary merger approximation. As shown below, the analytic formula we derive reproduces well the numerical results, therefore we conclude that for this matter, this approximation is not significant.

Without loss of generality, we may assume $dM\gg Mdx$, therefore the progenitor masses of mergers whose descendant mass is $M\pm dM/2$ and ratio is $x\pm dx/2$ are $M_1\pm dM_1/2\equiv\frac{x}{x+1}(M\pm dM/2)$ and $M_2\pm dM_2/2\equiv\frac{1}{x+1}(M\pm dM/2)$. Therefore, $r_d(z_d,M,x)dMdx$ is also exactly the number of mergers whose more massive progenitor mass is $M_1\pm dM_1/2$ and ratio is $x\pm dx/2$, as well as the number of mergers whose less massive progenitor mass is $M_2\pm dM_2/2$ and ratio is $x\pm dx/2$. Therefore we can write the equalities
\begin{eqnarray}
&r_d(z_d,M,x)dMdx=r_p(z_p,M_1,x)dM_1dx\nonumber\\
&r_d(z_d,M,x)dMdx=r_{p2}(z_p,M_2,x)dM_2dx.
\label{e:number_of_mergers}
\end{eqnarray}

The number of mergers {\it per descendant halo}, whose descendant mass is $M\pm dM/2$ and ratio is $x\pm dx/2$, is simply 
\begin{eqnarray}
R_d(z_d,M,x)dx=r_d(z_d,M,x)dMdx/N_h(z_d,M)dM,
\label{e:number_of_mergers_per_desc_halo}
\end{eqnarray}
and similarly the number of mergers {\it per progenitor halo}, whose more massive progenitor mass is $M_1$ and ratio is $x\pm dx/2$, is 
\begin{eqnarray}
&R_p(z_p,M_1,x)dx=r_p(z_p,M_1,x)dx/N_h(z_p,M_1)\nonumber\\
&=r_d(z_d,M,x)\frac{dM}{dM_1}dx/N_h(z_p,M_1)\nonumber\\
&=R_d(z_d,M,x)N_h(z_d,M)\frac{dM}{dM_1}dx/N_h(z_p,M_1),
\label{e:number_of_mergers_per_prog_halo_begin}
\end{eqnarray}
where equations (\ref{e:number_of_mergers}) and (\ref{e:number_of_mergers_per_desc_halo}) were used in the first and second equalities, respectively. Scaling the mass from $M_1$ to $M$, we finally arrive at
\begin{eqnarray}
R_p(z_p,M,x)=R_d(z_d,\frac{x+1}{x}M,x)\frac{x+1}{x}N_h(z_d,\frac{x+1}{x}M)/N_h(z_p,M),
\label{e:number_of_mergers_per_prog_halo}
\end{eqnarray}
which is the relation between the merger rate per more-massive-progenitor halo and the merger rate per descendant halo.

In an analogous way, the merger rate per less-massive-progenitor halo is
\begin{eqnarray}
R_{p2}(z_p,M,x)=R_d(z_d,(x+1)M,x)(x+1)N_h(z_d,(x+1)M)/N_h(z_p,M),
\label{e:number_of_mergers_per_small_prog_halo}
\end{eqnarray}
but in order to correspond to $x<1$ ratios as defined in \S\ref{s:rates} we define $R_p(z_p,M,x<1)dx=R_{p2}(z_p,M,\frac{1}{x})d\frac{1}{x}$, therefore
\begin{eqnarray}
R_p(z_p,M,x<1)=R_d(z_d,(\frac{1}{x}+1)M,\frac{1}{x})(\frac{1}{x}+1)N_h(z_d,(\frac{1}{x}+1)M)x^{-2}/N_h(z_p,M).
\label{e:number_of_mergers_per_prog_halo_small_x}
\end{eqnarray}

Equations (\ref{e:number_of_mergers_per_prog_halo}) \& (\ref{e:number_of_mergers_per_prog_halo_small_x}) construct the merger rate per progenitor halo as defined in \S\ref{s:rates} from the merger rate per descendant halo, and can be applied regardless of the source of the given merger rate per descendant halo.

In order to compare the two quantities, it is more convenient to examine equation (\ref{e:number_of_mergers_per_prog_halo}), because both definitions of the merger rate are quantified in its validity range, i.e.~$x>1$. The largest source of difference is the term $N_h(z_d,\frac{x+1}{x}M)/N_h(z_p,M)$. For $M$ smaller than the knee of the mass function, where it is roughly a power law with $M$, this term is constant with $M$ for a given $x$, i.e.~the trend of the merger rate with mass stays the same. As opposed to that, where the mass function begins to drop exponentially, this term decreases exponentially with $M$ as well, and therefore so does $R_p/R_d$.

For completeness we show also how the merger rate per descendant halo can be derived from the merger rate per progenitor halo:
\begin{eqnarray}
R_{d}(z_d,M,x)=R_p(z_p,(\frac{x}{x+1})M,x)\frac{x}{x+1}N_h(z_p,(\frac{x}{x+1})M)/N_h(z_d,M)\nonumber\\
=R_p(z_p,\frac{M}{x+1},\frac{1}{x})\frac{x^{-2}}{x+1}N_h(z_p,(\frac{M}{x+1}))/N_h(z_d,M),
\label{e:number_of_mergers_per_desc_halo_from_prog}
\end{eqnarray}
as well as how the two regimes of the merger rate per progenitor halo are related to one another:
\begin{eqnarray}
R_p(z_p,M,x)=R_p(z_p,\frac{M}{x},\frac{1}{x})x^{-3}N_h(z_p,\frac{M}{x})/N_h(z_p,M).
\label{e:number_of_mergers_per_prog_halo_small_and_large_relation}
\end{eqnarray}

\end{document}